\documentclass[aps, 10pt, twocolumn, pra, longbibliography, superscriptaddress]{revtex4-2}
\usepackage{graphicx}
\usepackage{xcolor}
\usepackage{hyperref}
\usepackage{orcidlink}

\usepackage[utf8]{inputenc}
\usepackage[T1]{fontenc}
\usepackage{mathptmx}
\usepackage{amsmath}
\usepackage{amssymb}
\usepackage{tabularx}

\begin{document}

\title{Stochastic dynamics from maximum entropy in action space}

\author{Fabricio Souza Luiz \orcidlink{0000-0002-6375-0939}}
\email{fsluiz@unicamp.br}
\affiliation{Instituto de F\'\i sica ``Gleb Wataghin'', Universidade Estadual de Campinas, Campinas, SP, Brazil}

\author{José Carlos Bellizotti Souza \orcidlink{0000-0001-8149-5575}}
\email{jcbsouza@dac.unicamp.br}
\affiliation{Instituto de F\'\i sica ``Gleb Wataghin'', Universidade Estadual de Campinas, Campinas, SP, Brazil}

\author{Lu\'{\i}sa Toledo Tude \orcidlink{0000-0002-4058-2274}}
\affiliation{Institute for Cross-Disciplinary Physics and Complex Systems (IFISC) UIB-CSIC, Campus Universitat Illes Balears, 07122, Palma de Mallorca, Spain.}

\author{Marcos C\'esar de Oliveira \orcidlink{0000-0003-2251-2632}}
\email{marcos@ifi.unicamp.br}
\affiliation{Instituto de F\'\i sica ``Gleb Wataghin'', Universidade Estadual de Campinas, Campinas, SP, Brazil}

\date{\today}

\begin{abstract}
We develop an information-theoretic formulation of stochastic dynamics in which the fundamental stochastic variable is the total action connecting spacetime points, rather than individual paths. By maximizing Shannon entropy over a joint distribution of actions and endpoints, subject to normalization and a constraint on the mean action, we obtain a Boltzmann-like distribution in action space. This framework reproduces the standard Brownian propagator in the nonrelativistic limit and naturally extends to relativistic regimes, where the Wiener construction fails to preserve Lorentz covariance. The approach bypasses functional integration over paths, makes the role of entropic degeneracy explicit through an action-space density of states, and provides a transparent connection between the principle of least action and statistical inference. We derive the density of states explicitly using large deviation theory, showing that it takes a Gaussian form centered at the minimal action, and rigorously justify the saddle-point approximation in the diffusive regime. The Markovian property of the resulting propagator is verified to hold via the Chapman--Kolmogorov equation, following from the additivity of the minimal action for free-particle dynamics. In the diffusive regime, the resulting dynamics are governed by a competition between extremization of the action and entropic effects, which can be interpreted in terms of an effective action free energy. Our results establish an unified, covariant, and information-based foundation for classical and relativistic stochastic processes.
\end{abstract}

\maketitle

\section{Introduction}
\label{sec:intro}

Information theory provides a quantitative framework to characterize uncertainty in physical systems independently of semantic content. In his seminal work, Shannon introduced a measure of information based solely on probabilistic structure, establishing entropy as a fundamental and operationally well-defined quantity~\cite{Shannon1948,Shannon1949}. By focusing on transmission, storage, and inference rather than meaning, Shannon’s formulation made information a precise object amenable to physical analysis.

Building on this foundation, Jaynes formulated the Maximum Entropy Principle (MaxEnt), which states that, given partial information about a system encoded in a set of constraints, the probability distribution that maximizes Shannon entropy represents the least biased inference consistent with that information~\cite{Jaynes1957a}. When applied to physics, MaxEnt naturally reproduces equilibrium statistical mechanics and provides a general framework for statistical inference that extends beyond equilibrium settings.

A dynamical extension of the Jaynes principle arises when the entropy maximization is performed over ensembles of trajectories rather than static states. This idea, later formalized under the name of Maximum Caliber, allows one to infer probability distributions over paths subject to dynamical constraints~\cite{Wang2004a,wang2005maximum,Wang2006a,Presse2013}. In this framework, one considers all possible trajectories connecting two spacetime points $a=(x_a,t_a)$ and $b=(x_b,t_b)$ within a time interval $\Delta t=t_b-t_a$, possibly discretized into intermediate configurations for calculational purposes. Each trajectory $k$ is assigned a conditional probability $p_k(b|a)$ describing the transition between the endpoints along that path.

Imposing normalization of probabilities together with a constraint on the expected action,
\begin{equation}
\langle A \rangle = \sum_k p_k(b|a)\, A_{ab}(k),
\end{equation}
Wang ~\cite{Wang2004a,wang2005maximum} has nicely demonstrated that maximizing the Shannon entropy,
$
S = -\sum_k p_k \ln p_k$,
leads to a path probability distribution of the form
\begin{equation}
p_k(b|a) = \frac{e^{-\eta A_{ab}(k)}}{Z(b,a)}, 
\qquad 
Z(b,a) = \sum_k e^{-\eta A_{ab}(k)},
\end{equation}
where $A_{ab}(k)$ is the classical action evaluated along trajectory $k$ connecting two position points $a$ and $b$, and $\eta$ is a Lagrange multiplier associated with the action constraint. In this statistical formulation, the most probable trajectory corresponds to the path that minimizes the action, providing an information-theoretic interpretation of the principle of stationary action. This approach has also been used to derive stochastic uncertainty relations \cite{wang2005non}.

Action-based formulations of stochastic processes have a longer history, predating Wang's entropy-maximization approach. The pioneering work of Onsager and Machlup~\cite{Onsager1953a} introduced a functional formalism assigning probabilities to individual Langevin trajectories via an action principle, providing the first rigorous connection between fluctuations and irreversible processes. Subsequent developments in large deviation theory, particularly the Freidlin--Wentzell framework~\cite{Freidlin1998}, established a systematic theory for calculating rare-event probabilities in stochastic systems through action functionals \footnote{At this point it is important to emphasize that these path-level approaches differ fundamentally from the one that shall be developed in the present work, which formulates the variational principle directly in \emph{action space}, marginalizing over trajectories to obtain probability distributions over total action values rather than assigning probabilities to individual paths.}.

When applied to diffusive processes such as Brownian motion, this construction reproduces well-known results of stochastic dynamics. By discretizing trajectories into time slices of duration $\Delta t$ and taking the continuum limit, the summation over paths becomes a functional integral that yields the Wiener measure~\cite{Wiener1923,Kac1949,Cameron1944}. In this limit, the Lagrange multiplier $\eta$ can be identified with $1/(2mD)$, where $m$ is the particle mass and $D$ is the diffusion coefficient, leading to the classical diffusion propagator. This connection makes explicit how stochastic dynamics can emerge from an entropic variational principle applied to path ensembles.

Extending this reasoning to relativistic stochastic processes, however, encounters fundamental difficulties. The standard Wiener construction relies on a partition of trajectories in configuration space with respect to a preferred time parameter, which is incompatible with Lorentz invariance. As a consequence, the resulting stochastic processes are not Lorentz covariant, spacelike fluctuations violate causality, and the use of proper time becomes ambiguous for non-timelike paths~\cite{Dudley1966a,Dunkel2009a}. These issues are not merely technical but reflect the fact that the Wiener measure itself is intrinsically nonrelativistic.

A notable resolution was independently proposed by Dunkel and Hänggi ~\cite{Dunkel2009a}, who observed that the nonrelativistic Wiener propagator can be rewritten in terms of the classical action. By replacing the nonrelativistic action with its relativistic counterpart, they obtained a manifestly Lorentz-covariant diffusion propagator. This approach yields a consistent relativistic generalization of Brownian motion, although the substitution of actions is introduced at a phenomenological level.

In this work, we show that Dunkel's relativistic diffusion kernel arises naturally from an information-theoretic variational principle when the inference problem is reformulated in action space rather than configuration space. Instead of maximizing entropy over distributions of spacetime trajectories, we maximize Shannon entropy over a distribution $p(A)$ of action values subject to appropriate constraints. This shift renders Lorentz covariance manifest at the level of the inferred probability distribution and avoids the ambiguities associated with spacetime discretization. In the nonrelativistic limit, our formulation reduces to Wang's path-entropy result, while in the relativistic regime it yields a covariant stochastic propagator without \emph{ad hoc} modifications. Our approach therefore provides a unified MaxEnt-based foundation for both classical and relativistic stochastic dynamics.

A central quantity in our formulation is the density of states $g(A,b)$, which quantifies the measure of trajectories arriving at endpoint $b$ with total action $A$. We derive this density explicitly for the free Brownian motion using large deviation theory, showing that it takes a Gaussian form centered at the minimal action with variance scaling as $\sigma_A^2 \sim m^2 D\,\Delta t$. This calculation rigorously justifies the saddle-point approximation employed in the diffusive regime and makes the framework fully self-contained. Furthermore, we demonstrate that the Markovian property of the resulting propagator emerges naturally as a consequence of the Gaussian kernel structure, verified explicitly through the Chapman--Kolmogorov equation, rather than being imposed as an a priori assumption. These technical results establish the action-space MaxEnt formulation as a rigorous alternative to path-based constructions.

This article is organized as follows. In Sec. II we introduce the maximum-entropy formulation in action space, defining the joint probability distribution over actions and endpoints and establishing its thermodynamic and information-theoretic structure. Section III applies this framework to stochastic dynamics, first recovering the classical Brownian propagator in the nonrelativistic regime and then deriving a manifestly Lorentz-covariant relativistic diffusion kernel. In Sec. IV we discuss the conceptual implications of the action-space formulation, its relation to standard path-based approaches, and its connections to fluctuation theorems and effective free-energy principles. Technical derivations and complementary examples are deferred to the Appendices.

\section{Methods}

\subsection{Maximum Entropy in Action Space}


\begin{figure*}[t]
\centering
\includegraphics[width=0.7\textwidth]{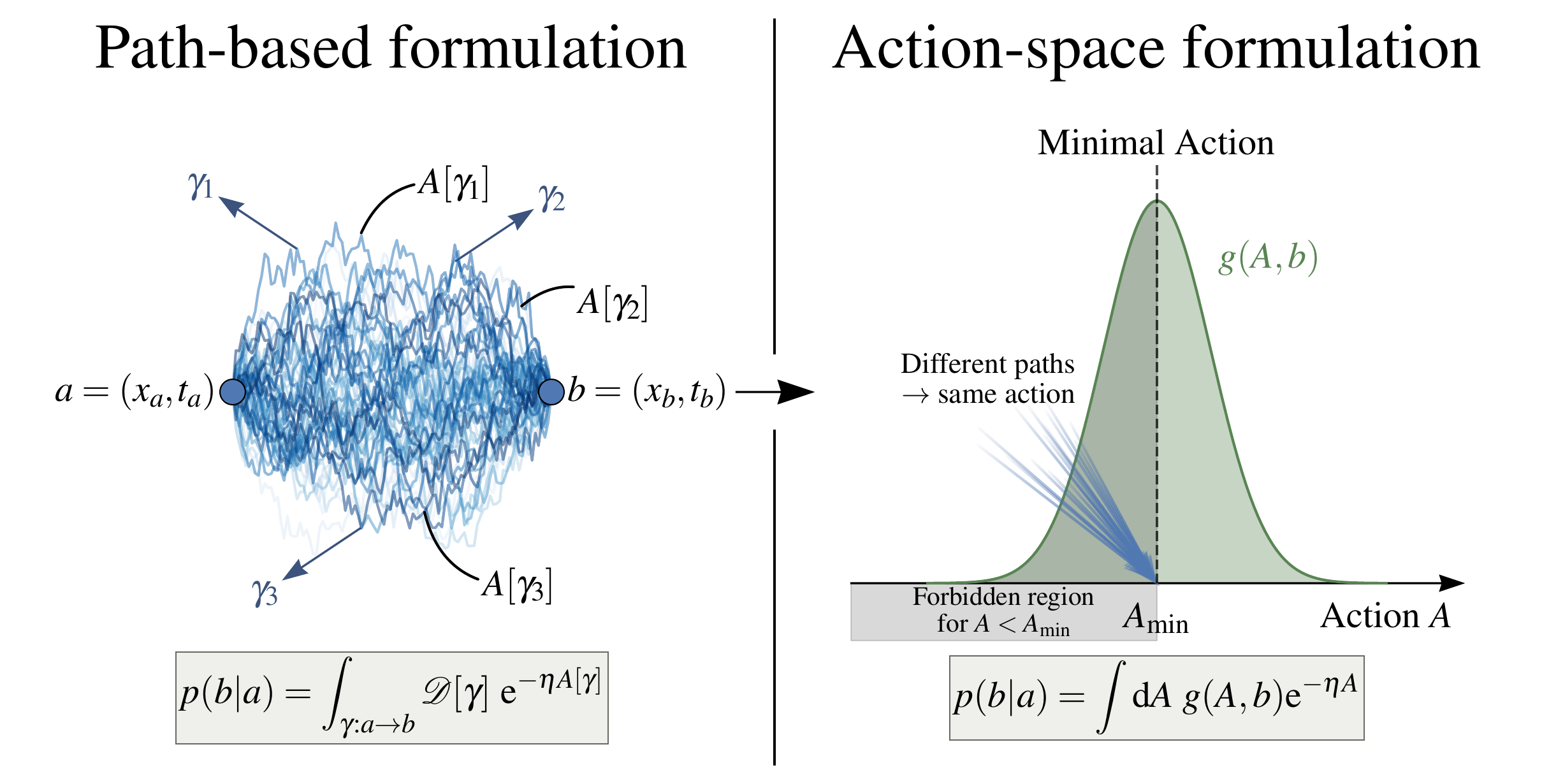}
\caption{
Schematic comparison between the path-based and action-space formulations of stochastic dynamics.
(\emph{Left}) In the conventional path-based approach, stochasticity is described by assigning probabilities to individual trajectories $\gamma$ connecting fixed spacetime endpoints $a=(x_a,t_a)$ and $b=(x_b,t_b)$, with the transition probability obtained from a functional integral over paths weighted by $e^{-\eta A[\gamma]}$. Entropic effects are implicit in the counting of trajectories and require discretization and a Wiener-type measure.
(\emph{Right}) In the action-space formulation proposed here, the total action $A$ is treated as the fundamental stochastic variable. Distinct paths leading to the same action value are grouped into a density of states $g(A,b)$, making path degeneracy explicit. The transition probability is obtained from an ordinary integral over action values, $p(b|a)=\int dA\, g(A,b)\,e^{-\eta A}$, with support restricted to $A\geq A_{\min}(b)$.
This reformulation bypasses functional integration, clarifies the role of entropy through $g(A,b)$, and renders Lorentz covariance manifest at the level of the probability distribution.
}
\label{fig:conceptual_scheme}
\end{figure*}

In the approach proposed by Wang in Refs.~\cite{Wang2004a,wang2005maximum,Wang2006a}, an ensemble of systems starting from an initial cell \(a\) evolves over a fixed time interval to reach various final cells \(b\) within a phase space volume \(B\). Each realization follows a specific trajectory \(k_b\) connecting \(a\) to \(b\), characterized by its action \(A_{k_b}\). The MaxEnt principle is applied to the distribution of paths, yielding probabilities \(p_{b|a}(k) \propto \exp[-\eta A_{ab}(k)]\), where the sum runs over all possible trajectories and all final destinations, as
\begin{equation}
\sum_{b \in B} \sum_{k} p_{b|a}(k) = 1.
\end{equation}
This construction naturally leads, in the continuum limit, to a functional integral (path integral) over trajectories, reproducing the Wiener measure for diffusive processes (see Appendix~\ref{app:wang}). Related dynamical entropy-based approaches are often discussed under the framework of Maximum Caliber, which generalizes Jaynes' MaxEnt principle to ensembles of trajectories~\cite{Presse2013}. The present formulation differs from these approaches by identifying the total action, rather than individual paths, as the fundamental stochastic variable.

We propose a fundamentally different formulation in which the stochastic variable is not the individual path but the \emph{total action} connecting two spacetime points. Instead of distinguishing between trajectories \(k_b\) that happen to have different actions, we treat the action \(A\) itself as the fundamental stochastic quantity. Multiple paths with the same total action \(A\) become indistinguishable at the level of the probability distribution, and their collective contribution is captured by a \emph{density of states} in action space. This reformulation, illustrated in Fig.~\ref{fig:conceptual_scheme}, shifts the inference problem from the space of individual trajectories to the space of total action values. Although the density of states in action space can be formally defined as the pushforward of a measure on path space, all probability assignments and variational principles are carried out directly in the reduced space $(A,b)$. As a result, the formalism avoids explicit functional integration at the level of entropy maximization and probability normalization, while retaining the full dynamical information encoded in the action.

Let \(p(b,A|a)\,dA\,d^dx_b\) denote the joint probability that, starting from spacetime point \(a=(x_a,t_a)\), the system arrives at point \(b=(x_b,t_b)\) within the infinitesimal spatial volume \(d^dx_b\) with total action in the interval \([A, A+dA]\), being the superscript $d$ representing the dimension of the system. Here, \(b\) varies over all spacetime points accessible from \(a\), and \(A\) varies over all physically admissible action values for each pair \((a,b)\).

Since different paths may connect \(a\) to \(b\) with the same total action, we introduce a \emph{density of states} \(g(A,b)\) that quantifies this degeneracy. Formally, \(g(A,b)\) is the pushforward of a measure on the space of admissible trajectories under the action functional. This construction is analogous to the contraction principle in large deviation theory~\cite{Freidlin1998,Touchette2009}, where distributions of macroscopic observables are induced from measures defined on microscopic configurations. For Brownian motion, \(g(A,b)\) can be derived microscopically from random collisions between the diffusing particle and thermal bath molecules, without assuming any specific stochastic equation (see Appendix~\ref{app:density_of_states}). The density of states is defined by
\begin{equation}
g(A,b) = \int \delta(A - A[\gamma])\,\delta^d(\mathbf{x}[\gamma](t_b) - \mathbf{x}_b)\,\mathcal{D}[\gamma],
\label{eq:density_of_states}
\end{equation}
where \(A[\gamma] = \int_{t_a}^{t_b} L(\gamma(t),\dot{\gamma}(t),t)\,dt\) is the action along path \(\gamma\), and the functional integral runs over all trajectories starting at \(a\). The quantity \(g(A,b)\,dA\,d^dx_b\) thus represents the measure (or "number") of trajectories arriving within \(d^dx_b\) around \(b\) with action between \(A\) and \(A+dA\).
Along this work we use the functional notation \(A[\gamma]\) to denote the action functional evaluated along a specific path \(\gamma\), and the scalar variable \(A\) (without arguments) to represent the total action as a real number. In probability distributions and integrals, \(A\) refers to the numerical value of the action, treated as a continuous stochastic variable.

Following Jaynes~\cite{Jaynes1957a}, in the absence of further information beyond the expected action, the least biased probability distribution maximizes the relative entropy~\cite{banavar2007a} between \(p(b,A|a)\) and the prior \(g(A,b)\):
\begin{equation}
S[p\|g]= -\int d^dx_b\int_{A_{\min}(a,b)}^{A_{\max}(a,b)} 
p(b,A|a)\,\ln\!\frac{p(b,A|a)}{g(A,b)}\,dA,
\label{eq:relative_entropy_action}
\end{equation}
subject to normalization and a constraint on the mean action:
\begin{align}
\int d^dx_b\int_{A_{\min}(b)}^{A_{\max}(b)}
p(b,A|a)\,dA &= 1,
\label{eq:action_norm}\\
\int d^dx_b\int_{A_{\min}(b)}^{A_{\max}(b)}
A\,p(b,A|a)\,dA &= \langle A\rangle.
\label{eq:action_mean}
\end{align}
Here, the integration over \(b\) runs over all spatial points in the final volume \(B\) at time \(t_b\), while for each pair \((a,b)\), the integration over \(A\) is restricted to physically admissible values. The limits \(A_{\min}(a,b)\) and \(A_{\max}(a,b)\) depend on the spacetime interval and physical constraints. For a free particle, \(A_{\min}(a,b)\) corresponds to the action along the classical (straight-line) trajectory connecting \(a\) to \(b\), while \(A_{\max}(a,b)\) may be infinite in the nonrelativistic case or finite in the relativistic case due to the speed-of-light constraint (see Sec.~\ref{sec:results} for explicit examples). 

Introducing Lagrange multipliers \(\alpha\) and \(\eta\) to enforce the constraints, we extremize the functional
\begin{align}
\mathcal{F}[p] = 
&- \iint p(b,A|a)\ln\!\frac{p(b,A|a)}{g(A,b)}\,dA\,d^dx_b \nonumber \\
&- \alpha\!\left(\iint p(b,A|a)\,dA\,d^dx_b - 1\right) \nonumber\\
&- \eta\!\left(\iint A\,p(b,A|a)\,dA\,d^dx_b - \langle A\rangle\right).
\end{align}
The extremization, \(\delta \mathcal{F}/\delta p = 0\) yields
\begin{equation}
-\ln\!\frac{p(b,A|a)}{g(A,b)} - 1 - \alpha - \eta A = 0,
\end{equation}
which, upon solving for \(p(b,A|a)\) and imposing the normalization constraint~\eqref{eq:action_norm}, gives
\begin{equation}
p(b,A|a) = \frac{g(A,b)\,e^{-\eta A}}{Z(\eta)},
\label{eq:pA_maxent}
\end{equation}
where the partition function is
\begin{equation}
Z(\eta) = \int d^dx_b\int_{A_{\min}(b)}^{A_{\max}(b)} g(A,b)\,e^{-\eta A}\,dA.
\label{eq:partition_function}
\end{equation}
We assume that the density of states \(g(A,b)\) grows at most sub-exponentially with the action \(A\), ensuring the convergence of the partition function \(Z(\eta)\) for \(\eta>0\). This condition is analogous to standard assumptions in equilibrium and nonequilibrium statistical mechanics and guarantees that the resulting probability distributions are normalizable.

Equation~\eqref{eq:pA_maxent} shows that the joint probability \(p(b,A|a)\) follows a Boltzmann distribution in the combined space of actions and spatial endpoints. The Lagrange multiplier \(\eta\) plays the role of an inverse weighting parameter in action space. We define the \emph{action-space information scale}\footnote{%
Despite the nomenclature ``informational temperature,'' the quantity \(T_{\text{info}} = 1/\eta\) has dimensions of action \([ML^2T^{-1}]\) rather than energy \([ML^2T^{-2}]\), and depends on system-specific parameters through the relation \(\eta = \gamma/(2k_BT)\), where \(\gamma\) is the friction coefficient. The term emphasizes a formal structural analogy with the inverse temperature \(\beta = 1/(k_BT)\) in equilibrium statistical mechanics: just as \(\beta\) controls exponential weighting \(e^{-\beta E}\) of energy states, \(\eta\) controls exponential weighting \(e^{-\eta A}\) of action values. This is a \emph{formal} analogy---\(T_{\text{info}}\) is not a thermodynamic temperature but rather a characteristic action scale of the system, proportional to the physical temperature through \(T_{\text{info}} = 2k_BT/\gamma = 2mk_BT\tau_{\text{relax}}\), where \(\tau_{\text{relax}} = m/\gamma\) is the relaxation time.%
} \(T_{\text{info}} \equiv 1/\eta\), which controls the relative weight of different action values. The exponential factor \(e^{-\eta A}\) favors low-action realizations, consistent with the principle of least action, while the density of states \(g(A,b)\) favors action values for which many distinct paths contribute.

The most probable outcome arises from a competition between these two effects, formally analogous to free energy minimization in thermodynamics. To make this analogy precise, we first define the \emph{reference density of states}:
\begin{equation}
g_0 \equiv g(A_{\min},b),
\label{eq:g0_definition}
\end{equation}
which anchors our effective potential at the classical (minimal) action. The \emph{effective potential} in action space is then:
\begin{equation}
\Phi(A,b) = A - T_{\text{info}} \ln\!\left[\frac{g(A,b)}{g_0}\right] = A - \frac{1}{\eta}\ln\!\left[\frac{g(A,b)}{g_0}\right].
\label{eq:effective_potential}
\end{equation}
By construction, \(\Phi(A_{\min},b) = A_{\min}\), ensuring that the effective potential is anchored at the classical action value, and differences \(\Delta\Phi = \Phi(A,b) - \Phi(A_{\min},b)\) are independent of \(g_0\). The probability distribution then takes the Boltzmann-like form \(p(A,b|a) \propto e^{-\eta \Phi(A,b)}\). It is important to note that \(\Phi(A,b)\) is a property of a specific microstate \((A,b)\), not the thermodynamic free energy of the ensemble.

The true (informational)  action free ``energy'' of the ensemble is defined as
\begin{equation}
F(\eta) = -\frac{1}{\eta} \ln Z(\eta) = -T_{\text{info}} \ln Z(\eta),
\label{eq:free_energy_true}
\end{equation}
which satisfies the fundamental thermodynamic relation
\begin{equation}
F = \langle A \rangle - T_{\text{info}} S,
\label{eq:helmholtz_action}
\end{equation}
where the mean action is \(\langle A \rangle = -\partial \ln Z / \partial \eta\) and the ensemble entropy is
\begin{equation}
S = \ln Z(\eta) + \eta \langle A \rangle.
\label{eq:entropy_action}
\end{equation}
This parallels the equilibrium thermodynamic relations \(F = U - TS\) and \(S = k_B(\ln Z + \beta U)\), with \(A\) playing the role of energy \(U\) and \(\eta\) playing the role of inverse temperature \(\beta\). It shows an intrinsic relation of the minimization of the  action free energy of the ensemble with the minimization of the expected action and the maximization of the ensemble entropy.  It is important to remark that this correspondence is structural only, not dimensional. The analogy captures the mathematical form of the variational principle and the competition between cost minimization and entropy maximization, but \(T_{\text{info}}\) has dimensions of action \([ML^2T^{-1}]\) rather than temperature \([ML^2T^{-2}]\), and depends on system-specific parameters (see Table~\ref{tab:action_thermo}).

The system preferentially realizes action values that minimize \(\Phi(A,b)\) rather than the action itself.  Even if the density of states \(g(A,b)\) is sharply peaked at some \(A_* > A_{\min}(b)\), the value \(A_*\) may be preferred over the minimal action provided
\begin{equation}
A_* - A_{\min} < T_{\text{info}} \ln\left[\frac{g(A_*,b)}{g(A_{\min},b)}\right].
\label{eq:entropic_compensation}
\end{equation}
This condition states that the entropic gain, weighted by the informational temperature, can compensate for the excess action cost. Conversely, the classical path \(A_{\min}\) is preferred when the inverse condition holds:
\begin{equation}
A_* - A_{\min} > T_{\text{info}} \ln\left[\frac{g(A_*,b)}{g(A_{\min},b)}\right].
\label{eq:no_compensation}
\end{equation}
This inverted condition reveals two distinct physical regimes. When \(g(A_*,b) > g(A_{\min},b)\) (entropy favors fluctuations), Eq.~\eqref{eq:no_compensation} indicates that the action cost \(\Delta A\) exceeds the entropic benefit \(T_{\text{info}} \ln[g(A_*,b)/g(A_{\min},b)]\), preventing compensation. More fundamentally, when \(g(A_*,b) < g(A_{\min},b)\) (entropy \emph{also} favors the classical path), the logarithmic term becomes negative, and Eq.~\eqref{eq:no_compensation} is automatically satisfied for any \(\Delta A > 0\). In this regime, the system faces a double penalty for deviating from \(A_{\min}\): both the action cost \emph{and} the entropic cost \(T_{\text{info}} |\ln[g(A_*,b)/g(A_{\min},b)]|\) oppose fluctuations, reinforcing the principle of least action. Table~\ref{tab:entropic_regimes} summarizes these two contrasting regimes.

\begin{table*}[t]
\centering
\caption{Comparison of entropic compensation regimes for systems with density of states \(g(A,b)\) peaked either away from or at the minimal action \(A_{\min}\). The ratio \(R \equiv g(A_*,b)/g(A_{\min},b)\) characterizes the entropic bias.}
\label{tab:entropic_regimes}
\begin{tabular}{lcc}
\hline\hline
\textbf{Property} & \(g(A_*,b) > g(A_{\min},b)\) & \(g(A_*,b) < g(A_{\min},b)\) \\
 & (Entropy favors fluctuations) & (Entropy favors minimum) \\
\hline
\(\ln[g(A_*,b)/g(A_{\min},b)]\) & \(> 0\) (entropic gain) & \(< 0\) (entropic loss) \\
Entropic barrier & Potential benefit & Additional cost \\
Compensation condition & \(\Delta A < T_{\text{info}} \ln R\) & Impossible \\
Inverted condition & \(\Delta A > T_{\text{info}} \ln R\) & Always satisfied \\
\(\Phi(A_{\min})\) vs \(\Phi(A_*)\) & Depends on \(T_{\text{info}}\) & \(\Phi(A_{\min}) < \Phi(A_*)\) always \\
Critical temperature & \(T_{\text{crit}} = \Delta A / \ln R\) & Formally defined\(^*\) \\
Effect of increasing \(T_{\text{info}}\) & Favors compensation & Increases barrier \\
Dominant mechanism & Competition & Mutual reinforcement \\
Physical examples & Multi-well potentials, & Brownian motion, \\
 & instantons & simple diffusion \\
\hline\hline
\multicolumn{3}{l}{\footnotesize \(^*\)When \(R < 1\), \(\ln R < 0\) yields \(T_{\text{crit}} < 0\), which is unphysical. No positive temperature can induce entropic compensation in this regime.}
\end{tabular}
\end{table*}

For the free Brownian particle studied in Sec.~\ref{sec:results}, the density of states is Gaussian-peaked precisely at \(A_{\min}(b)\), so \(g(A_*,b) < g(A_{\min},b)\) for all \(A_* \neq A_{\min}\). Consequently, Eq.~\eqref{eq:entropic_compensation} is never satisfied, and the minimal action remains the most probable. However, this mechanism becomes relevant for systems with non-Gaussian action distributions, where \(g(A,b)\) can exhibit peaks away from \(A_{\min}\). For instance, in double-well potentials \cite{instanton1,instanton2}, instanton trajectories that tunnel between metastable states have higher action than direct paths but acquire large statistical weight \(g(A_*)\) due to temporal freedom in the barrier-crossing event: the instanton can occur at any time within the interval \([t_a, t_b]\), generating a continuum of distinct paths with the same action \(A_*\). When the entropic gain \(T_{\text{info}} \ln[g(A_*)/g(A_{\min})]\) exceeds the action cost \(\Delta A = A_* - A_{\min}\), the system preferentially realizes fluctuating paths over the classical minimum-action trajectory. A discrete toy model illustrating this competition is provided in Appendix~\ref{app:toy_model}, and detailed discussion of physical examples (multi-well potentials, activated processes, relativistic causality constraints) is given in Appendix~\ref{app:density_of_states}, Subsection~6.

This compensation mechanism differs fundamentally from the least action principle applied to deterministic trajectories -- here, the most probable \emph{action} balances dynamical optimization with statistical degeneracy, and the temperature \(T_{\text{info}}\) sets the scale at which entropic effects become comparable to action differences.

Physically, the action \(A\) can be interpreted as a macroscopic observable characterizing the ``cost'' of a transition, while \(g(A,b)\) encodes the microscopic multiplicity of paths realizing that cost. The distribution~\eqref{eq:pA_maxent} thus provides a statistical interpretation of action functionals, analogous to how microcanonical ensembles relate energy to entropy in equilibrium statistical mechanics.

From the joint distribution \(p(b,A|a)\), we can extract two physically relevant marginal probabilities. The first one, the probability of arriving at a point $b$ by integration over all feasible actions,
\begin{eqnarray}
p(b|a) &=& \int_{A_{\min}(b)}^{A_{\max}(b)} p(b,A|a)\,dA\nonumber\\ &=& \frac{1}{Z(\eta)}\int_{A_{\min}(b)}^{A_{\max}(b)} g(A,b)\,e^{-\eta A}\,dA.
\label{eq:marginal_b}
\end{eqnarray}
This is the standard propagator describing the transition probability from \(a\) to \(b\), analogous to the one obtained by Wang through summation over paths (see Appendix~\ref{app:wang}). Conversely, the probability of realizing a specific action value \(A\), integrated over all final spatial positions, is
\begin{equation}
p(A|a) = \int d^dx_b\, p(b,A|a) = \frac{1}{Z(\eta)}\int d^dx_b\, g(A,b)\,e^{-\eta A}.
\label{eq:marginal_A}
\end{equation}
While \(p(b|a)\) has direct physical interpretation as a transition amplitude, \(p(A|a)\) quantifies the distribution of action values realized by the ensemble of trajectories emanating from \(a\). In stochastic processes, \(p(A|a)\) provides insight into the fluctuations of the total "cost" of transitions, complementary to the spatial distribution \(p(b|a)\).

Although both Wang's formulation and ours sum over all possible final destinations \(b\), they differ fundamentally in the nature of the stochastic variable and the structure of the probability space.
In Wang's approach, the fundamental stochastic variable is the \emph{individual path} \(k\) connecting \(a\) to \(b\). Each trajectory is treated as a distinct entity, characterized by its action \(A_{ab}(k)\). The MaxEnt distribution assigns probability
\begin{equation}
p_{b|a}(k) = \frac{e^{-\eta A_{ab}(k)}}{Z}, \quad Z = \sum_{b \in B} \sum_{k} e^{-\eta A_{ab}(k)},
\end{equation}
where the double sum runs over all final cells \(b\) and all paths \(k\) reaching them. In the continuum limit, this sum over discrete paths becomes a functional integral,
\begin{equation}
p(b|a) \propto \int_{\gamma: a \to b} \mathcal{D}[\gamma]\, e^{-\eta S[\gamma]},
\end{equation}
which, for diffusive processes, yields the Wiener measure. This construction naturally imposes a Markovian structure when paths are discretized into infinitesimal steps -- the transition probability between successive points depends only on the current state, not on the history.

In our approach, the fundamental stochastic variable is the \emph{action value} \(A\), treated as a continuous scalar quantity. Different paths with the same total action are indistinguishable at the level of the probability distribution, their collective contribution is encoded in the density of states \(g(A,b)\). The MaxEnt distribution is defined over the space \((b,A)\), not over individual paths as in Eqs.~\eqref{eq:pA_maxent},\eqref{eq:partition_function}. The transition probability is obtained by integrating over actions as shown in Eq.~\eqref{eq:marginal_b}, which is an ordinary integral, not a functional integral.

Mathematically, the two formulations are related by a change of variables. Wang's functional integral can be rewritten using~\eqref{eq:density_of_states} 
\begin{equation}
\int_{\gamma: a \to b} \mathcal{D}[\gamma]\, e^{-\eta S[\gamma]} = \int dA\, g(A,b)\, e^{-\eta A},
\end{equation}
where \(g(A,b)\) acts as the Jacobian of the transformation from path space to action space. However, the conceptual and computational advantages of the action-space formulation are substantial and we enumerate it here:

\textit{(i) Bypassing of functional integration.} Functional integrals are notoriously difficult to define rigorously and compute explicitly. By integrating directly over action values, we replace an infinite-dimensional integration with a one-dimensional integral over \(A\), greatly simplifying calculations.

 \textit{(ii) No path discretization required.} Wang's approach requires partitioning trajectories into \(N\) intermediate steps and taking the limit \(N \to \infty\). This discretization introduces a Markovian structure at the level of path increments. In contrast, the action \(A\) is a global quantity depending on the entire trajectory. No intermediate steps are needed, and no Markovian assumption is imposed at the level of path discretization or intermediate time slicing. Markovian behavior, when present, emerges only in specific physical limits, such as short-time or weak-correlation regimes, rather than being built into the formalism at the outset.
 
\textit{(iii) Manifest Lorentz covariance.} The action \(A = \int L\,dt\) is a Lorentz scalar for relativistic systems. A probability distribution defined over action values therefore transforms covariantly under Lorentz transformations. While in the present formulation the final event \(b\) is specified on a fixed time hypersurface for concreteness, the covariance of the weighting factor \(e^{-\eta A}\) is independent of any particular foliation of spacetime.
In contrast, the Wiener measure and path-based constructions rely on foliating spacetime into spacelike hypersurfaces (constant-time slices), which breaks manifest covariance. This difference becomes crucial in relativistic contexts, where proper-time parametrization and causality constraints complicate path-based formulations.

\textit{(iv) Explicit degeneracy accounting.} The density \(g(A,b)\) explicitly quantifies how many paths contribute to each action value, making the role of entropic effects transparent. In Wang's formulation, this degeneracy is implicit in the sum over paths and only becomes apparent in the continuum limit.

\textit{(v) Generality.} The formulation requires only a well-defined action functional and a corresponding measure on the space of admissible trajectories, encoded through \(g(A,b)\). No assumptions about equilibrium, near-equilibrium conditions, or specific stochastic forces are imposed at the level of the variational principle. This makes the framework applicable to a broad class of systems, including relativistic and potentially quantum-relativistic regimes.

While both approaches yield the same transition probabilities when properly evaluated, the action-space formulation provides a more direct, computationally tractable, and conceptually transparent route to stochastic dynamics. It separates the dynamical principle (minimization of action via \(e^{-\eta A}\)) from the statistical principle (maximization of entropy via \(g(A,b)\)), making the interplay between deterministic and stochastic effects explicit.

In the deterministic limit (\(\eta \to \infty\)), the Boltzmann factor \(e^{-\eta A}\) becomes sharply peaked around the minimum action \(A_{\min}\). If \(g(A,b)\) varies slowly near \(A_{\min}\), the integral~\eqref{eq:marginal_b} is dominated by trajectories with \(A \approx A_{\min}\), recovering the principle of least action. However, for finite \(\eta\), the distribution balances action minimization with entropic considerations encoded in \(g(A,b)\). This provides a natural interpolation between deterministic mechanics (\(\eta \to \infty\), action dominates) and purely entropic dynamics (\(\eta \to 0\), density of states dominates).

In the next subsection, we apply this formalism to classical Brownian motion, recovering Wang's results from the action-space perspective and establishing an explicit connection to the diffusion equation.

\subsection{Application to Brownian Motion: Theoretical Framework}

To illustrate the consistency and physical meaning of the MaxEnt formulation in action space, we apply it to classical and relativistic Brownian motion. These examples are particularly instructive as Wang originally derived the nonrelativistic Brownian propagator using a path-based MaxEnt approach~\cite{Wang2004a,Wang2006a}, providing a direct benchmark for comparison. Moreover, the relativistic case demonstrates how the action-space formulation naturally preserves Lorentz covariance, overcoming well-known difficulties of the Wiener measure in relativistic regimes~\cite{Dunkel2009a,Dudley1966a}.

In both cases, we consider a free particle of mass \(m\) (or rest mass \(m_0\) in the relativistic case) undergoing stochastic motion characterized by a diffusion coefficient \(D\). Between successive collisions with the surrounding medium, the particle follows free-particle dynamics~\cite{Wang2006a,Dunkel2007a,schilling2014brownian}. Our goal is to derive the transition probability \(p(b|a)\) from spacetime point \(a=(x_a,t_a)\) to \(b=(x_b,t_b)\) over a time interval \(\Delta t = t_b - t_a\).

Within our MaxEnt formulation, the density \(g(A,b)\) quantifies the measure of trajectories arriving at spatial position \(x_b\) at time \(t_b\) with total action \(A\). By definition, \(g(A,b) = 0\) for \(A < A_{\min}(b)\), since no trajectory connecting the endpoints can have action below the classical minimum. For \(A \geq A_{\min}(b)\), the density counts the number of distinct paths realizing the given action value.

For a free Brownian particle, the density of states takes a Gaussian form centered at \(A_{\min}(b)\),
\begin{equation}
g(A,b) \propto \exp\left[-\frac{(A - A_{\min}(b))^2}{2\sigma_A^2}\right], \quad A \geq A_{\min}(b).
\label{eq:gaussian_density}
\end{equation}
This form arises because action fluctuations \(\Delta A = A - A_{\min}\) result from the accumulation of many small, independent velocity increments induced by thermal noise. By the central limit theorem, the distribution of \(\Delta A\) approaches a Gaussian in the diffusive regime \(\Delta t \gg \tau_{\text{relax}}\) (see Appendix~\ref{app:density_of_states} for a rigorous derivation). The Gaussian is truncated at \(A = A_{\min}\), but for typical parameters the probability mass below this cutoff is exponentially small (\(\lesssim 10^{-10}\)) and contributes negligibly to physical observables.

For particles in external potentials \(V(x,\lambda)\), the density of states \(g(A,b;\lambda)\) depends on the potential parameters and may exhibit peaks away from \(A_{\min}\), for instance, in double-well potentials where instanton trajectories acquire large statistical weight. Such generalizations are discussed in Appendix~\ref{app:density_of_states}, Subsection~6.

Fluctuations of the total action arise from stochastic velocity increments induced by thermal noise. For Gaussian diffusion governed by a Langevin process, the variance of the accumulated action over a time interval \(\Delta t\) scales as \footnote{This scaling can be derived microscopically from random collisions with thermal bath molecules, without assuming a specific stochastic equation. For a free Brownian particle, uncorrelated velocity kicks from Maxwell-Boltzmann distributed collisions yield $\sigma_A^2 \sim m k_B T D\,\Delta t$ via the central limit theorem in the diffusive regime $\Delta t \gg \tau_{\text{relax}}$ (see Appendix~\ref{app:density_of_states}, Section 1). The result can also be verified using the Langevin equation, giving $\sigma_A^2 \sim m^2 D^2\,\Delta t/\tau_{\text{relax}}$, which is equivalent under the Einstein relation (Appendix~\ref{app:density_of_states}, Section 2).}
\[
\sigma_A^2 \sim m k_B T D\,\Delta t,
\]
reflecting the quadratic growth of velocity fluctuations and their time integration along the trajectory~\cite{Wang2006a,Onsager1953a,Zwanzig2001}. The density \(g(A,b)\) is therefore appreciably different from zero over a range \(\Delta A \sim \sigma_A = \sqrt{m k_B T D \Delta t}\) around the minimal action \(A_{\min}(b)\).

The transition probability \(p(b|a)\) is obtained from Eq.~\eqref{eq:marginal_b} by integrating \(g(A,b)\,e^{-\eta A}\) over \(A\). The exponential factor \(e^{-\eta A}\) introduces a characteristic decay scale \(1/\eta\). To assess whether \(g(A,b)\) varies slowly compared to the exponential, we compare these two scales. The logarithmic derivative of the exponential is
\begin{equation}
\left|\frac{d\ln e^{-\eta A}}{dA}\right| = \eta,
\end{equation}
while for \(g(A,b)\), the characteristic scale is
\begin{equation}
\left|\frac{d\ln g(A,b)}{dA}\right| \sim \frac{1}{\sigma_A} = \frac{1}{\sqrt{m^2 D \Delta t}}.
\end{equation}

When the observation time satisfies \(\Delta t \gg \tau_{\text{relax}}\), where \(\tau_{\text{relax}} \sim m/\gamma\) is the microscopic relaxation time associated with friction coefficient \(\gamma\), the dimensionless parameter \(\eta \sigma_A\) becomes large,
\begin{equation}
\eta \sigma_A \sim \eta \sqrt{m^2 D \Delta t}
= \frac{1}{2mD} \sqrt{m^2 D \Delta t}
= \frac{1}{2}\sqrt{\frac{\Delta t}{\tau_{\text{relax}}}}
\gg 1,
\end{equation}
where we used \(D/m = k_BT/(m^2\gamma) = \tau_{\text{relax}} k_BT/m\). This condition is satisfied when \(\Delta t \gg 4\tau_{\text{relax}}\). For a colloidal particle in water (mass \(m \sim 10^{-15}\) kg, friction coefficient \(\gamma \sim 10^{6}\) s\(^{-1}\), giving \(\tau_{\text{relax}} \sim 10^{-9}\) s), the approximation is excellent for observation times \(\Delta t \gtrsim 10^{-8}\) s, which encompasses virtually all experimentally relevant timescales. For \(\Delta t = 100\tau_{\text{relax}}\), we have \(\eta\sigma_A \approx 5\), and the relative error in the saddle-point approximation is \(\lesssim 1\%\). The approximation breaks down when \(\Delta t \sim \tau_{\text{relax}}\), corresponding to the ballistic regime where inertial effects dominate and the Gaussian form of \(g(A,b)\) is no longer valid.

The condition \(\eta\sigma_A \gg 1\) implies that the exponential factor varies much more rapidly than \(g(A,b)\) in the integration region. In this regime, corresponding to the diffusive limit where \(\Delta t\) exceeds microscopic relaxation times \(\tau_\text{relax}\), 
\(g(A,b)\) can be approximated by its value at \(A_{\min}(b)\) (for a rigorous derivation and discussion of the regime of validity, see Appendix~\ref{app:density_of_states}, Subsection~4),
\begin{equation}
p(b|a) \approx \frac{g(A_{\min}(b),b)}{Z(\eta)} \int_{A_{\min}(b)}^{A_{\max}(b)} e^{-\eta A}\, dA.
\label{eq:p_ba_slowly_varying}
\end{equation}
Under this approximation, the spatial dependence of the propagator is determined entirely by the minimal action \(A_{\min}(b)\), while \(g(A_{\min}(b),b)\) contributes only to the overall normalization. This approximation is formally equivalent to a Laplace (saddle-point) approximation of the action integral and is well justified in the diffusive regime, where fluctuations around the minimal action are small compared to the exponential decay scale~\cite{Bender1978a}.

Evaluating the integral in Eq.~\eqref{eq:p_ba_slowly_varying}, we obtain the general form of the transition probability,
\begin{equation}
p(b|a) = \frac{g(A_{\min}(b),b)}{Z(\eta)} \left[e^{-\eta A_{\min}(b)} - e^{-\eta A_{\max}(b)}\right].
\label{eq:pab_general_form}
\end{equation}
The normalization constant \(Z(\eta)\) is determined by the condition \(\int p(b|a)\, d^dx_b = 1\), and absorbs all factors independent of \(x_b\), including \(g(A_{\min}(b),b)\) if it varies slowly with \(b\). In the nonrelativistic case, \(A_{\max} \to +\infty\) and the term \(e^{-\eta A_{\max}}\) vanishes. In the relativistic case, causality restricts admissible trajectories to timelike or lightlike paths, which in turn bounds the action from above for fixed endpoints. As a result, the upper limit \(A_{\max}\) is finite and determined by the light-cone structure of spacetime~\cite{Dudley1966a,Dunkel2009a}.

In the following subsections, we apply Eq.~\eqref{eq:pab_general_form} to derive the propagators for classical and relativistic Brownian motion.

\section{Results}
\label{sec:results}
\subsection{Classical Brownian Motion}

We now demonstrate that the action-space MaxEnt formulation reproduces the standard Brownian propagator. For a free particle of mass \(m\) moving in one spatial dimension between points \(x_a\) and \(x_b\) over a time interval \(\Delta t = t_b - t_a\), the classical action is
\begin{equation}
A[x(t)] = \int_{t_a}^{t_b} L\, dt = \int_{t_a}^{t_b} \frac{1}{2} m \dot{x}^2 \, dt.
\label{eq:classical_action_functional}
\end{equation}

To find the minimal action, we extremize \(A[x(t)]\) subject to the boundary conditions \(x(t_a) = x_a\) and \(x(t_b) = x_b\). The Euler-Lagrange equation yields constant velocity \(\dot{x} = (x_b - x_a)/\Delta t\), giving
\begin{equation}
A_{\min}(b) = \frac{m(x_b - x_a)^2}{2\Delta t}.
\label{eq:classical_action_min}
\end{equation}

For the nonrelativistic case, there is no upper limit on the particle's velocity. Trajectories can exhibit arbitrarily large fluctuations, resulting in arbitrarily large actions. Therefore,
\begin{equation}
A_{\max} = +\infty.
\label{eq:classical_action_max}
\end{equation}

Substituting Eqs.~\eqref{eq:classical_action_min} and~\eqref{eq:classical_action_max} into Eq.~\eqref{eq:pab_general_form}, and noting that in the limit \(A_{\max} \to +\infty\) the term \(e^{-\eta A_{\max}} \to 0\) is exponentially suppressed, we obtain
\begin{equation}
p(b|a) = \mathcal{N}\, e^{-\eta A_{\min}(b)} = \mathcal{N}\, \exp\left[-\eta\,\frac{m(x_b - x_a)^2}{2\Delta t}\right],
\label{eq:pba_exp_action}
\end{equation}
where \(\mathcal{N}\) is the normalization constant determined by the condition
\begin{equation}
\int_{-\infty}^{+\infty} p(b|a)\, dx_b = 1.
\end{equation}

To evaluate \(\mathcal{N}\), we compute the Gaussian integral:
\begin{equation}
1 = \mathcal{N} \int_{-\infty}^{+\infty} \exp\left[-\eta\,\frac{m(x_b - x_a)^2}{2\Delta t}\right] dx_b = \mathcal{N} \sqrt{\frac{2\pi \Delta t}{\eta m}},
\end{equation}
yielding
\begin{equation}
\mathcal{N} = \sqrt{\frac{\eta m}{2\pi \Delta t}}.
\end{equation}
Thus, the normalized propagator reads
\begin{equation}
p(b|a) = \sqrt{\frac{\eta m}{2\pi \Delta t}}\,\exp\left[-\eta\,\frac{m(x_b - x_a)^2}{2\Delta t}\right].
\label{eq:gaussian_diffusion_eta}
\end{equation}

To connect \(\eta\) with the diffusion coefficient \(D\), we compute the second moment of the distribution. Using the standard formula for the variance of a Gaussian with exponent \(-\alpha(x-\mu)^2\), where \(\langle (x-\mu)^2 \rangle = 1/(2\alpha)\), we have
\begin{equation}    
\langle(x_b - x_a)^2\rangle = \int_{-\infty}^{+\infty} (x_b - x_a)^2\, p(b|a)\, dx_b\nonumber = \frac{\Delta t}{\eta m}.
\end{equation}

For standard Brownian motion, the mean-square displacement is given by Einstein's relation
\begin{equation}
\langle(x_b - x_a)^2\rangle = 2D\Delta t,
\end{equation}
where \(D\) is the diffusion coefficient. Equating the two expressions, we obtain
\begin{equation}
\frac{\Delta t}{\eta m} = 2D\Delta t \quad \Rightarrow \quad \eta = \frac{1}{2mD}.
\label{eq:eta_identification_classical}
\end{equation}

Substituting Eq.~\eqref{eq:eta_identification_classical} into Eq.~\eqref{eq:gaussian_diffusion_eta}, we recover the standard diffusion kernel:
\begin{equation}
p(b|a) = \frac{1}{\sqrt{4\pi D \Delta t}}\,\exp\left[-\frac{(x_b - x_a)^2}{4 D \Delta t}\right].
\label{eq:diffusion_kernel}
\end{equation}

Having derived the propagator~\eqref{eq:diffusion_kernel} without explicitly imposing Markovian structure at the variational level, we now verify \emph{a posteriori} that it satisfies the Chapman--Kolmogorov equation, confirming consistency with Markovian dynamics. For any intermediate time \(t_c\) with \(t_a < t_c < t_b\), the Chapman--Kolmogorov relation states 
\begin{equation}
p(x_b, t_b | x_a, t_a) = \int_{-\infty}^{+\infty} p(x_b, t_b | x_c, t_c)\, p(x_c, t_c | x_a, t_a)\, dx_c.
\label{eq:chapman_kolmogorov}
\end{equation}

To verify this, we compute the convolution of two Gaussian kernels. Let \(\tau_1 = t_c - t_a\) and \(\tau_2 = t_b - t_c\), so that \(\tau_1 + \tau_2 = t_b - t_a\). Substituting the propagator~\eqref{eq:diffusion_kernel}we obtain
\begin{align}
&\int_{-\infty}^{+\infty} \frac{1}{\sqrt{4\pi D \tau_2}}\exp\left[-\frac{(x_b - x_c)^2}{4D\tau_2}\right] \nonumber \\
&\quad \times \frac{1}{\sqrt{4\pi D \tau_1}}\exp\left[-\frac{(x_c - x_a)^2}{4D\tau_1}\right] dx_c.
\label{eq:convolution_integral}
\end{align}

This is the convolution of two Gaussian distributions with means \(x_a\) and \(x_c\) and variances \(2D\tau_1\) and \(2D\tau_2\), respectively. By the standard result for Gaussian convolution, the integral evaluates to a Gaussian with mean \(x_a\) (when propagating from \(x_a\) to \(x_c\)) and variance \(2D(\tau_1 + \tau_2) = 2D(t_c - t_a)\),
\begin{align}
\int_{-\infty}^{+\infty} p(x_b|x_c;\tau_2)\, & p(x_c|x_a;\tau_1)\, dx_c = \nonumber\\
&\frac{1}{\sqrt{4\pi D(t_b - t_a)}}\exp\left[-\frac{(x_b - x_a)^2}{4D(t_b - t_a)}\right],
\label{eq:markov_verified}
\end{align}
which is precisely \(p(x_b, t_b | x_a, t_a)\) as given by Eq.~\eqref{eq:diffusion_kernel} with \(\Delta t = t_b - t_a\). Thus, the propagator derived from the action-space MaxEnt formulation satisfies the Chapman--Kolmogorov equation, confirming consistency with Markovian dynamics. While the variational principle itself does not explicitly impose this structure, the Markovian property is inherited from the underlying microscopic assumption of independent, uncorrelated collisions with the thermal bath (see Appendix~\ref{app:density_of_states}) and preserved through the coarse-graining to action space. Conversely, Markovianity can fail if the thermal bath exhibits memory effects, such that noise correlations decay on timescales comparable to the observation interval: \(\langle \xi(t)\xi(t')\rangle \neq \delta(t-t')\) with correlation time \(\tau_{\text{bath}} \sim \Delta t\). In such non-Markovian regimes, the density of states \(g(A,b)\) would acquire dependence on the trajectory history, requiring generalization beyond the Gaussian form derived here. Similarly, anomalous diffusion processes (\(\langle x^2 \rangle \propto t^\alpha\), with \(\alpha \neq 1\)) or systems with long-range temporal correlations would necessitate modified forms of \(g(A,b)\) to capture non-local effects in action space.


The diffusion coefficient \(D\) is related to the temperature \(T\) and friction coefficient \(\gamma\) via Einstein's relation
\begin{equation}
D = \frac{k_B T}{m\gamma},
\end{equation}
where \(k_B\) is Boltzmann's constant. Substituting into Eq.~\eqref{eq:eta_identification_classical}, we find
\begin{equation}
\eta = \frac{1}{2mD} = \frac{\gamma}{2k_B T},
\end{equation}
where we used \(mD = k_B T/\gamma\) from the Einstein relation. This establishes a direct connection between the Lagrange multiplier \(\eta\) and the inverse temperature, weighted by the friction coefficient \(\gamma\).

\emph{Note on conventions.} We define \(\gamma\) as the inverse relaxation time, \(\tau_{\text{relax}} = m/\gamma\), with dimensions \([T^{-1}]\). The damping force in the Langevin equation is \(F_{\text{drag}} = -m\gamma \dot{x}\), giving the equation of motion
\begin{equation}
m\ddot{x} = -m\gamma \dot{x} + \xi(t),
\end{equation}
where \(\xi(t)\) is Gaussian white noise satisfying \(\langle \xi(t)\xi(t')\rangle = 2m\gamma k_B T \delta(t - t')\). Einstein's relation \(D = k_B T/(m\gamma)\) follows from the fluctuation-dissipation theorem, balancing thermal fluctuations against viscous damping. This convention differs from the Stokes friction coefficient \(\gamma_{\text{Stokes}}\) (with dimensions \([MT^{-1}]\)), which appears in the relation \(D = k_B T/\gamma_{\text{Stokes}}\) for a macroscopic particle in a viscous fluid. Our convention is standard in the literature of Langevin equation ~\cite{risken1989fokker}, where \(\gamma\) represents a damping rate (relaxation frequency) rather than a viscous drag coefficient.

Equation~\eqref{eq:diffusion_kernel} is precisely the Brownian propagator derived by Wang using a path-based MaxEnt approach (see Eq. (29) in Ref.~\cite{Wang2006a}) and coincides with the fundamental solution of the Fokker–Planck equation. This agreement confirms the consistency of the action-space formulation. Importantly, our derivation did not require explicit path decomposition, infinitesimal time slicing, or the imposition of a Markovian structure. The Markovian property of the resulting propagator arises \emph{a posteriori} from the Gaussian form of the kernel and its convolution properties, rather than being imposed at the level of the variational principle. The diffusive dynamics emerge naturally from maximizing Shannon entropy in the space of actions, subject to constraints on normalization and mean action. In this sense, the action-space MaxEnt formalism provides a coarse-grained representation of stochastic dynamics, where the statistics of the total action encode all relevant information about the diffusion process.

\subsection{Relativistic Brownian Motion}

The extension of stochastic dynamics to the relativistic domain presents well-known conceptual and mathematical challenges. The standard Wiener process, which underlies nonrelativistic diffusion, is defined in position space with a privileged time parameter and assumes independent Gaussian increments. This formulation is not Lorentz invariant and can lead to superluminal propagation when naively applied to relativistic systems~\cite{Dunkel2009a,Dudley1966a}. As a result, a variety of alternative formulations of relativistic diffusion have been proposed~\cite{Hakim1968a,debbasch2004diffusion, Dunkel2005a, Dunkel2009a}. However, attempts to construct Lorentz-invariant diffusion equations often rely on \emph{ad hoc} modifications or phase-space constraints that obscure the underlying variational structure.

In contrast, the MaxEnt formulation in action space provides a natural route to relativistically consistent stochastic dynamics. Because the action is a Lorentz scalar, probability distributions defined over actions are automatically covariant under Lorentz transformations. The uncertainty concerns the distribution of total actions \(A\) connecting two spacetime points, rather than the paths themselves, eliminating the need for a specific time parametrization and avoiding explicit breaking of Lorentz symmetry.

For a free relativistic particle of rest mass \(m_0\) moving in one spatial dimension, the action between spacetime points \(a=(x_a,t_a)\) and \(b=(x_b,t_b)\) is
\begin{equation}
A[x(t)] = -m_0 c^2 \int_{t_a}^{t_b} \sqrt{1 - \frac{\dot{x}^2}{c^2}}\, dt,
\label{eq:relativistic_action_functional}
\end{equation}
where \(c\) is the speed of light.\footnote{We adopt the convention \(A = -m_0 c \int ds\) (negative action for timelike paths) so that \(A_{\min} < 0\) for subluminal motion and \(A_{\max} = 0\) at the light cone. This choice ensures that the action is bounded above by zero, with the causal boundary (lightlike trajectories) serving as a natural upper limit. The negative sign convention is consistent with the structure \(A \in [A_{\min}(b), 0]\) used throughout the formalism, where \(A_{\min}(b)\) corresponds to the classical (minimum-action) trajectory and \(A_{\max} = 0\) reflects the relativistic causal constraint. This sign choice is natural for the MaxEnt formulation: since the Boltzmann weight \(e^{-\eta A}\) with \(\eta > 0\) favors \emph{smaller} (more negative) actions, the classical trajectory with \(A_{\min} < 0\) is exponentially preferred over near-lightlike fluctuations with \(A \to 0\). The alternative convention \(A = +m_0 c \int ds > 0\) would require \(\eta < 0\) to maintain the same physical behavior, complicating the thermodynamic analogy.} Let \(\Delta t = t_b - t_a\) and \(\Delta x = x_b - x_a\). Extremizing the action subject to the boundary conditions \(x(t_a) = x_a\) and \(x(t_b) = x_b\) yields a constant velocity \(\dot{x} = \Delta x / \Delta t\), giving the minimal action
\begin{equation}
A_{\min}(b) = -m_0 c^2 \Delta t \sqrt{1 - \frac{(\Delta x)^2}{c^2 (\Delta t)^2}} = -m_0 c \sqrt{c^2 (\Delta t)^2 - (\Delta x)^2}.
\label{eq:min_action_relativistic}
\end{equation}

Causality imposes the constraint \(|\Delta x| \leq c\Delta t\), ensuring that the spacetime interval is timelike or lightlike. The minimal action is realized for subluminal motion, with \(A_{\min}\) becoming more negative as the particle moves slower (further from the light cone). 

For the maximal action, we note that as the particle's velocity approaches \(c\), the integrand in Eq.~\eqref{eq:relativistic_action_functional} vanishes. A particle moving at the speed of light (\(\dot{x} = \pm c\)) yields
\begin{equation}
A_{\max} = 0.
\label{eq:max_action_relativistic}
\end{equation}
Although a massive particle cannot follow a null trajectory in a deterministic sense, lightlike paths naturally appear here as limiting configurations of admissible stochastic trajectories. In the present formulation, the upper bound \(A_{\max}=0\) reflects a causal constraint on the support of the probability distribution rather than a physical equation of motion, and does not imply that individual realizations propagate exactly at the speed of light. This corresponds to a null trajectory along the light cone. The causal constraint thus restricts the action to the interval \(A \in [A_{\min}(b), 0]\), with \(A_{\min}(b) < 0\) for timelike separations.

Substituting Eqs.~\eqref{eq:min_action_relativistic} and~\eqref{eq:max_action_relativistic} into Eq.~\eqref{eq:pab_general_form}, we obtain the transition probability
\begin{equation}
p(b|a) = \frac{1}{\mathcal{N}}\left[\exp\left(\eta m_0 c\sqrt{c^2(\Delta t)^2 - (\Delta x)^2}\right) - 1\right],
\label{eq:relativistic_propagator_unnorm}
\end{equation}
defined for \(|\Delta x| \leq c\Delta t\). The normalization constant \(\mathcal{N}\) is determined by
\begin{equation}
\int_{x_a - c\Delta t}^{x_a + c\Delta t} p(b|a)\, dx_b = 1.
\end{equation}

Introducing the dimensionless variable \(u = \Delta x / (c\Delta t) \in [-1, 1]\) and the parameter \(\alpha = \eta m_0 c \Delta t\), the normalization integral becomes
\begin{equation}
\mathcal{N} = c\Delta t \int_{-1}^{1} \left[e^{\alpha\sqrt{1-u^2}} - 1\right] du.
\end{equation}

This integral (calculated explicitly in Appendix \ref{app:norm_relat_k}) can be expressed in terms of modified Bessel functions \(I_n\) and modified Struve functions \(L_n\), as
\begin{equation}
\mathcal{N} = c\Delta t\, \left[ \pi\big(I_1(\alpha)+L_{-1}(\alpha)\big) - 2 \right].
\label{eq:normalization_relativistic}
\end{equation}
where \(I_1(\alpha)\) is the modified Bessel function of the first kind and \(L_{-1}(\alpha)\) is the modified Struve function. Thus, the normalized relativistic propagator is
\begin{equation}
p(b|a) = \frac{1}{\mathcal{N}}\left[\exp\left(\alpha\sqrt{1 - \frac{(\Delta x)^2}{c^2(\Delta t)^2}}\right) - 1\right].
\label{eq:relativistic_propagator}
\end{equation}

To verify consistency with the classical result, we expand Eq.~\eqref{eq:relativistic_propagator} in the nonrelativistic limit \(|\Delta x| \ll c\Delta t\), corresponding to \(|u| \ll 1\). Expanding the square root in Eq.~\eqref{eq:min_action_relativistic},
\begin{equation}
\sqrt{1 - \frac{(\Delta x)^2}{c^2(\Delta t)^2}} = 1 - \frac{(\Delta x)^2}{2c^2(\Delta t)^2} + \mathcal{O}\left(\frac{v^4}{c^4}\right),
\end{equation}
we obtain
\begin{eqnarray}
A_{\min} && \approx -m_0 c^2 \Delta t \left[1 - \frac{(\Delta x)^2}{2c^2(\Delta t)^2}\right] = -m_0 c^2 \Delta t + \frac{m_0 (\Delta x)^2}{2\Delta t} \nonumber \\&&+ \mathcal{O}(v^4/c^4).
\end{eqnarray}

The first term \(-m_0 c^2 \Delta t\) is a constant rest-energy contribution independent of \(\Delta x\). The second term reproduces the nonrelativistic kinetic action, identifying \(m_0\) with the nonrelativistic mass \(m\). The exponential in Eq.~\eqref{eq:relativistic_propagator_unnorm} becomes
\begin{equation}
\exp\left(\eta m_0 c\sqrt{c^2(\Delta t)^2 - (\Delta x)^2}\right) \approx e^{\eta m_0 c^2 \Delta t}\,\exp\left[-\eta\,\frac{m_0(\Delta x)^2}{2\Delta t}\right].
\end{equation}

The leading constant factor \(e^{\eta m_0 c^2 \Delta t} \gg 1\) dominates over the \(-1\) in Eq.~\eqref{eq:relativistic_propagator_unnorm}, and cancels in the normalization. The resulting distribution is Gaussian:
\begin{equation}
p(b|a) \approx \frac{1}{\tilde{\mathcal{N}}}\,\exp\left[-\eta\,\frac{m_0(\Delta x)^2}{2\Delta t}\right],
\end{equation}
which has the same form as Eq.~\eqref{eq:pba_exp_action} with \(m = m_0\).

To identify \(\eta\), we compute the variance in this limit. For the Gaussian distribution above, we have
\begin{equation}
\langle(\Delta x)^2\rangle = \frac{\Delta t}{\eta m_0}.
\end{equation}
Comparing with the Einstein relation \(\langle(\Delta x)^2\rangle = 2D\Delta t\), we obtain
\begin{equation}
\eta = \frac{1}{2m_0 D} .
\label{eq:eta_identification_relativistic}
\end{equation}

In natural units where \(c=1\), consistent with the nonrelativistic identification. Using the Einstein relation \(D = k_B T/\gamma\), we find
\begin{equation}
\eta = \frac{\gamma }{2m_0 k_B T},
\end{equation}
confirming that \(\eta\) is inversely proportional to the thermal energy \(k_B T\), weighted by the friction coefficient and the speed of light. This demonstrates that the relativistic formulation, based on constraining the action to the causal interval \([A_{\min}, 0]\), coherently reduces to standard Brownian motion in the nonrelativistic limit.

The relativistic propagator~\eqref{eq:relativistic_propagator} coincides with the result obtained by Dunkel et al.~\cite{Dunkel2009a}, who constructed a Lorentz-covariant diffusion equation by replacing the nonrelativistic action in the Wiener kernel with its relativistic counterpart. Our derivation shows that this substitution is not \emph{ad hoc} but emerges naturally from the MaxEnt principle applied in action space. By formulating the uncertainty in terms of action values rather than spacetime paths, manifest Lorentz covariance is automatically preserved, and the causal constraint \(|\Delta x| \leq c\Delta t\) is naturally incorporated through the bounds on \(A\). This provides a principled, information-theoretic foundation for relativistic stochastic processes, generalizing Wang's path-based formulation to regimes where the Wiener measure fails.

\section{Discussion}

We have developed a MaxEnt formulation of stochastic dynamics in which the fundamental stochastic variable is the total action connecting spacetime points, rather than the paths themselves. By maximizing Shannon entropy over a distribution \(p(b,A|a)\) of action values—subject to normalization and a constraint on the mean action—we derived a Boltzmann-like distribution that reproduces known results for classical Brownian motion while naturally extending to relativistic regimes where the standard Wiener construction fails.

The key insight is that uncertainty in paths can be effectively represented as uncertainty in actions, with the density of states $g(A,b)$ encoding the degeneracy of trajectories with the same total action.
This reformulation provides manifest Lorentz covariance, eliminates the need for explicit path discretization at the level of the variational principle, and establishes a direct connection between variational principles (least action) and statistical inference (maximum entropy).


Wang's path-based formulation assigns probability to individual trajectories, requiring discretization into infinitesimal steps and naturally imposing a Markovian structure. In contrast, the action-space formulation treats the total action as a global observable, bypassing the need for path decomposition. This shift from local to global description has several conceptual advantages as enumerated below:

\begin{enumerate}
\item \textit{No explicit Markovian decomposition.} The action \(A\) is a global observable that depends on the entire trajectory, avoiding the need to decompose paths into infinitesimal Markovian steps. While the underlying microscopic dynamics assume a Markovian bath (independent collisions), the action-space formulation does not require this structure to be imposed explicitly at the variational level. The Chapman--Kolmogorov equation is verified \emph{a posteriori} as a consistency check rather than as a foundational assumption.

\item \textit{Manifest covariance.} Since the action is a Lorentz scalar, probability distributions over actions are automatically covariant. This eliminates the difficulties associated with defining path measures on spacetime foliations, which break manifest Lorentz symmetry.

\item \textit{Physical transparency.} The competition between action minimization (dynamical principle) and entropic degeneracy (statistical principle) is explicit in the distribution \(p(A,b) \propto g(A,b) e^{-\eta A}\). This makes clear that the most probable transitions balance deterministic optimization with statistical multiplicity.
\end{enumerate}

From a practical standpoint, the action-space formulation replaces functional integration over infinite-dimensional path spaces with ordinary integration over action values. While the density of states \(g(A,b)\) must be calculated or approximated, this is often simpler than evaluating path integrals directly. In the saddle-point regime (\(\eta \gg 1/\sigma_A\)), the propagator reduces to \(p(b|a) \propto e^{-\eta A_{\min}(b)}\), requiring only the classical action -- a significant computational advantage for systems where \(A_{\min}\) is analytically tractable but the full path integral is not.


The formal structure of the action-space MaxEnt distribution closely parallels equilibrium statistical mechanics. To make this connection explicit, the correspondence between equilibrium thermodynamics and the action-space formulation is summarized in Table~\ref{tab:action_thermo}.
\begin{table*}[t]
\caption{Formal structural analogy between equilibrium thermodynamics and the action-space formulation of stochastic dynamics. The correspondence is mathematical rather than dimensional: \(T_{\text{info}} = 1/\eta\) has dimensions of action \([ML^2T^{-1}]\), not temperature \([ML^2T^{-2}]\), and the analogy highlights structural similarities in the variational principles. The table emphasizes that both frameworks minimize an effective free energy through competition between a cost term (energy \(E\) or action \(A\)) and an entropic term (\(TS\) or \(T_{\text{info}}\ln g\)).}
\label{tab:action_thermo}
\begin{ruledtabular}
\begin{tabular}{ll}
\textbf{Equilibrium Thermodynamics} & \textbf{Action-Space Dynamics} \\
Energy $E$ & Action $A$ \\
Density of states $\Omega(E)$ & Density of states $g(A,b)$ \\
Boltzmann entropy $S = k_B \ln \Omega(E)$ & Path entropy $S_{\text{path}} = k_B \ln g(A,b)$ \\
Temperature $T$ & Informational temperature $T_{\text{info}} = 1/\eta$ \\
Inverse temperature $\beta = 1/(k_B T)$ & Lagrange multiplier $\eta = 1/T_{\text{info}}$ \\
Canonical distribution $p(E) \propto \Omega(E) e^{-\beta E}$
& Action distribution $p(A,b) \propto g(A,b) e^{-\eta A}$ \\
Helmholtz free energy $F = U - TS$
& Free energy $F = \langle A \rangle - T_{\text{info}} S$ \\
Microstate energy $E$ & Microstate action $A$ \\
Effective potential & Effective potential $\Phi(A,b) = A - T_{\text{info}} \ln[g(A,b)/g_0]$ \\
Entropy $S = k_B(\ln Z + \beta U)$ & Entropy $S = \ln Z + \eta \langle A \rangle$ \\
\end{tabular}
\end{ruledtabular}
\end{table*}

This analogy suggests that stochastic dynamics can be understood as minimizing an \emph{effective potential} (cf. Eq.~\eqref{eq:effective_potential}). In the saddle-point approximation, where the dominant contribution comes from \(A \approx A_{\min}(b)\), this reduces to
\begin{equation}
\Phi(b|a) = A_{\min}(b) - T_{\text{info}} \ln g(A_{\min}(b), b),
\end{equation}
which balances the deterministic cost (minimal action) against the entropic gain (logarithm of the density of states), weighted by the informational temperature. In the saddle-point regime $\eta\sigma_{A} \gg 1$, the effective potential is dominated by contributions near $A_{\min}$. The system does not simply minimize action but rather minimizes this effective potential, explaining why trajectories deviate from the classical path when \(g(A,b)\) varies significantly.

For Brownian motion, the identification \(\eta = 1/(2mD) = \gamma/(2k_B T)\) establishes that the action-space parameter \(T_{\text{info}} = 1/\eta = 2mD = 2k_BT/\gamma\) sets a characteristic action scale proportional to the physical temperature \(T\). This proportionality reflects the fact that thermal energy \(k_BT\) drives action fluctuations via the diffusion coefficient \(D = k_BT/(m\gamma)\), which in turn determines the width of the action distribution: \(\sigma_A \sim \sqrt{mk_BTD\,\Delta t} = \sqrt{2mD\,\Delta t} \cdot \sqrt{k_BT/2} \sim \sqrt{T_{\text{info}} \cdot k_BT \cdot \Delta t/(2D)}\). The parameter \(\eta = 1/T_{\text{info}}\) sets the relative weight of action values in the exponential factor \(e^{-\eta A}\). In the limit \(\eta \to \infty\) (corresponding to \(T \to 0\) or strong friction \(\gamma \to \infty\)), the weighting becomes sharply peaked, recovering deterministic least-action dynamics. Conversely, \(\eta \to 0\) (corresponding to \(T \to \infty\) or weak friction \(\gamma \to 0\)) yields broad, entropy-dominated distributions where action cost becomes negligible compared to statistical multiplicity.

\subsection{Fluctuation Theorems and Thermodynamics in Action Space}
\label{sec:fluctuation_theorems}

The thermodynamic structure of the action-space formulation admits a natural connection to fluctuation theorems. Since the fundamental statistical weight is \(e^{-\eta A[\gamma]}\) with \(\eta\) having dimensions of inverse action, the natural conjugate quantity in fluctuation relations is \emph{action}, not energy. This leads to an ``action-work'' formulation that parallels the standard Jarzynski and Crooks relations ~\cite{Jarzynski1997,Crooks1999,landi2021irreversible}.

\subsubsection{Micropath ensemble}

At the microscopic level, each path \(\gamma: a \to b\) has action \(A[\gamma]\) and probability
\begin{equation}
\mathbb{P}(\gamma|a) = \frac{e^{-\eta A[\gamma]}}{Z(\eta)}, \qquad Z(\eta) = \sum_{\gamma: a \to \star} e^{-\eta A[\gamma]}.
\label{eq:micropath_probability}
\end{equation}
The coarse-grained distribution, \(p(A,b|a) = g(A,b)\,e^{-\eta A}/Z(\eta)\), arises from marginalizing over paths with the same action value, where \(g(A,b)\) counts the degeneracy.

\subsubsection{Action-work definition}

To derive fluctuation theorems, we compare two path ensembles: a ``forward'' process with protocol \(\lambda_F(t)\) and its ``reverse'' \(\lambda_R(t) = \lambda_F(\tau - t)\). These define two action functionals,
\begin{equation}
A_F[\gamma] = A[\gamma; \lambda_F], \qquad A_R[\tilde{\gamma}] = A[\tilde{\gamma}; \lambda_R],
\label{eq:forward_reverse_action}
\end{equation}
where \(\tilde{\gamma}\) denotes the time-reversed path. The \emph{action-work} is defined as the variation of action between forward and reverse experiments,
\begin{equation}
\mathcal{W}_A[\gamma] \equiv A_F[\gamma] - A_R[\tilde{\gamma}].
\label{eq:action_work}
\end{equation}
This quantity has dimensions of action \([ML^2T^{-1}]\), with conjugate parameter \(\eta\) (dimensions \([ML^2T^{-1}]^{-1}\)), ensuring that \(\eta \mathcal{W}_A\) is dimensionless.

\subsubsection{Crooks-like relation}

The path measures for forward and reverse processes are
\begin{equation}
\mathbb{P}_F[\gamma] = \frac{e^{-\eta A_F[\gamma]}}{Z_F(\eta)}, \qquad \mathbb{P}_R[\tilde{\gamma}] = \frac{e^{-\eta A_R[\tilde{\gamma}]}}{Z_R(\eta)}.
\label{eq:path_measures}
\end{equation}
Taking the ratio:
\begin{align}
\frac{\mathbb{P}_F[\gamma]}{\mathbb{P}_R[\tilde{\gamma}]} &= \exp\!\big(-\eta[A_F[\gamma] - A_R[\tilde{\gamma}]\big) \cdot \frac{Z_R}{Z_F} \nonumber\\
&= \exp\!\big(-\eta\,\mathcal{W}_A[\gamma]\big) \cdot \frac{Z_R}{Z_F}.
\label{eq:ratio_paths}
\end{align}
Defining the \emph{free action} (analogous to free energy):
\begin{equation}
{\mathcal{F}(\eta) \equiv -\eta^{-1} \ln Z(\eta), \qquad \Delta\mathcal{F} \equiv \mathcal{F}_R - \mathcal{F}_F = -\eta^{-1} \ln\frac{Z_R}{Z_F}},
\label{eq:free_action}
\end{equation}
the ratio becomes the \emph{Crooks fluctuation theorem in action space}:
\begin{equation}
{\ln \frac{\mathbb{P}_F[\gamma]}{\mathbb{P}_R[\tilde{\gamma}]} = -\eta\big(\mathcal{W}_A[\gamma] - \Delta\mathcal{F}\big)}.
\label{eq:crooks_action_space}
\end{equation}
This has the same structure as the standard Crooks relation, with \(\eta\) replacing \(\beta\) and action-work replacing mechanical work.

\subsubsection{Jarzynski-like equality}

Summing over all paths weighted by the forward measure,
\begin{align}
\langle e^{-\eta \mathcal{W}_A} \rangle_F &= \sum_\gamma \mathbb{P}_F[\gamma]\, e^{-\eta(A_F[\gamma] - A_R[\tilde{\gamma}])} \nonumber\\
&= \frac{1}{Z_F} \sum_\gamma e^{-\eta A_R[\tilde{\gamma}]} = \frac{Z_R}{Z_F}.
\label{eq:jarzynski_derivation}
\end{align}
This yields the \emph{Jarzynski equality in action space},
\begin{equation}
{\langle e^{-\eta \mathcal{W}_A} \rangle_F = \frac{Z_R(\eta)}{Z_F(\eta)} = e^{-\eta \Delta\mathcal{F}}}.
\label{eq:jarzynski_action_space}
\end{equation}

\subsubsection{Second law in action space}

By Jensen's inequality, \(\langle e^{-\eta \mathcal{W}_A} \rangle \geq e^{-\eta \langle \mathcal{W}_A \rangle}\). Combined with Eq.~\eqref{eq:jarzynski_action_space},
\begin{equation}
{\langle \mathcal{W}_A \rangle \geq \Delta\mathcal{F}}.
\label{eq:second_law_action_space}
\end{equation}
This is the second law of thermodynamics in action space -- the average action-work required to change the control parameter is at least as large as the free action difference, with equality for reversible (quasi-static) processes. The dissipated action-work \(\mathcal{W}_{\text{diss}} = \langle \mathcal{W}_A \rangle - \Delta\mathcal{F} \geq 0\) quantifies irreversibility.

\subsubsection{Terminology}

We emphasize that \(\mathcal{W}_A\) is an \emph{action-work}, not mechanical work in the conventional sense. The term reflects the structural analogy with thermodynamics: just as \(\beta W\) is dimensionless in standard fluctuation theorems (with \(\beta = 1/(k_BT)\) and \(W\) in energy units), here \(\eta \mathcal{W}_A\) is dimensionless (with \(\eta\) in inverse-action units and \(\mathcal{W}_A\) in action units). The standard energetic Jarzynski equality \(\langle e^{-\beta W} \rangle = e^{-\beta \Delta F}\) can be recovered by introducing time-averaged quantities: since \(H = -\partial A/\partial t\) from the Hamilton-Jacobi equation, the mechanical work over a process is related to the action-work via \(W = -\Delta(\partial A/\partial t) = \Delta H\). For quasi-static processes where the time derivative can be evaluated, this provides an explicit map between the action-space and energy-space formulations.

\subsubsection{Interpretation for free particles}

For the free Brownian particle studied in this work, where no external protocol \(\lambda(t)\) is applied, the fluctuation theorems take a simplified form. In the absence of time-dependent driving, the forward and reverse processes are statistically equivalent: \(A_F(\gamma) = A_R(\tilde{\gamma})\) for corresponding paths, implying \(\mathcal{W}_A = 0\) and \(\Delta\mathcal{F} = 0\). The Jarzynski equality becomes trivially \(\langle e^0 \rangle = 1\), and the Crooks relation reduces to microscopic reversibility \(\mathbb{P}_F(\gamma) = \mathbb{P}_R(\tilde{\gamma})\).

The fluctuation theorems become nontrivial when the system is driven by a time-dependent protocol---for instance, a particle in a potential \(V(x,\lambda(t))\) where \(\lambda\) is externally controlled. In such cases, the action functional depends explicitly on \(\lambda(t)\), and the action-work \(\mathcal{W}_A = A[\gamma;\lambda_F] - A[\tilde{\gamma};\lambda_R]\) quantifies the asymmetry between forward and reverse processes. The free-particle results derived in Secs.~\ref{sec:results} and~\ref{sec:fluctuation_theorems} thus provide the foundation for extensions to driven systems, where the full power of the action-space fluctuation theorems becomes manifest.


The action-space formulation is mathematically equivalent to Wang's path-based approach when the density \(g(A,b)\) correctly encodes the path measure. However, the two perspectives offer complementary insights. Path integrals emphasize the sum-over-histories interpretation of quantum and stochastic mechanics, while the action-space view highlights the competition between optimization (minimal action) and statistics (maximal entropy). Both recover the same propagators, but the action-space formulation makes manifest the role of Lorentz covariance and avoids the technical difficulties of defining functional measures on constrained path spaces.

Our formulation also bears conceptual similarity to Nelson's stochastic mechanics, where quantum dynamics arise from a diffusion process in configuration space~\cite{nelson1966derivation}. While Nelson postulates forward and backward diffusion to recover Schrödinger's equation, our approach derives stochastic dynamics from information-theoretic principles without reference to quantum mechanics. Nevertheless, the shared emphasis on action and diffusion suggests potential connections that merit further exploration, particularly in the semiclassical regime.


The derivations presented here rely on the approximation that \(g(A,b)\) varies slowly compared to \(e^{-\eta A}\), justified when \(\eta \sigma_A \gg 1\). This condition holds in the diffusive regime \(\Delta t \gg \tau_\text{relax}\), where thermal equilibration has occurred. At very short times or in systems with rapid transients, this approximation breaks down, and the full dependence on \(g(A,b)\) must be retained. In principle, \(g(A,b)\) can be calculated from the underlying path integral, but this may be technically challenging for complex systems. Developing systematic approximation schemes for \(g(A,b)\)—analogous to semiclassical or mean-field methods in quantum mechanics—remains an important direction for future work.

The formulation presented here is entirely classical, relying on classical actions and probability distributions. Extension to quantum systems, where amplitudes replace probabilities and interference effects become important, is nontrivial. In the semiclassical limit, one might expect a hybrid formulation where \(g(A,b)\) encodes quantum fluctuations around classical trajectories. However, a fully quantum version would require a path-integral formulation with complex weights, where the role of entropy maximization is less clear. Whether an information-theoretic principle analogous to MaxEnt governs quantum dynamics is a deep open question in the foundations of quantum mechanics.

\section{Conclusion}
The reformulation of stochastic dynamics in action space demonstrates that fundamental physical principles—variational optimization and information-theoretic inference—are two sides of the same coin. By shifting the uncertainty from paths to actions at the level of stochastic inference, we gain computational simplicity, manifest covariance, and a transparent connection to thermodynamics. The results presented here establish a unified, covariant, and information-based foundation for free-particle stochastic dynamics across classical and relativistic regimes, as explicitly demonstrated for Brownian motion. Extending the present framework to interacting systems, external potentials, and quantum processes, as well as connections to martingale theory~\cite{manzano2022survival}, remains an important direction for future research.
While these extensions present significant conceptual and technical challenges, the free-particle framework developed here provides a solid starting point and demonstrates the viability of action-space maximum entropy formulations beyond the nonrelativistic domain.

\begin{acknowledgments}

M.C.O. acknowledges partial financial support from the National Institute of Science and Technology for Applied Quantum
Computing through CNPq (Process No.~408884/2024-0) and from the São Paulo Research Foundation (FAPESP), through the Center
for Research and Innovation on Smart and Quantum Materials (CRISQuaM, Process No.~2013/07276-1).
\end{acknowledgments}
\appendix
\section{Action  and Jaynes MaxEnt principle}
\label{app:wang}

In Refs.~\cite{Wang2004a,Wang2006a}, Wang formulated a variational approach to stochastic dynamics based on the 
\emph{Maximum Entropy Principle} (MaxEnt) applied to ensembles of trajectories. 
Consider a set of paths \(k\) connecting two spacetime points \(a=(x_a,t_a)\) and \(b=(x_b,t_b)\), separated by a time interval 
\(\Delta t = t_b - t_a\). 
Each path \(k\) corresponds to a sequence of intermediate states 
\(\{x_1,x_2,\ldots,x_{N-1}\}\) and is assigned a transition probability \(p_k(b|a)\).
The uncertainty over the actual path taken is quantified by the \emph{path Shannon entropy}:
\begin{equation}
H_{ab} = - \sum_b\sum_k p_k(b|a)\ln p_k(b|a).
\label{eq:entropy_path}
\end{equation}

Maximizing \(H_{ab}\) subject to normalization and to the known average action,
\begin{align}
\sum_b\sum_k p_k(b|a) &= 1, \label{eq:const_norm}\\
\sum_b\sum_k p_k(b|a)\,A_{ab}(k) &= \langle A_{ab}\rangle, \label{eq:const_action}
\end{align}
yields, via Lagrange multipliers \(\alpha\) and \(\eta\), the functional
\begin{eqnarray}
\mathcal{F} &=& -\sum_b\sum_k p_k \ln p_k- \alpha\left(\sum_b\sum_k p_k - 1\right)\nonumber \\&&
- \eta \left(\sum_b\sum_k p_k A_{ab}(k) - \langle A_{ab}\rangle\right),
\end{eqnarray}
whose extremization leads to a Boltzmann-type distribution in path space,
\begin{equation}
p_k(b|a) = \frac{1}{Z_{ab}}\,e^{-\eta A_{ab}(k)}, 
\qquad 
Z_{ab} = \sum_b\sum_k e^{-\eta A_{ab}(k)}.
\label{eq:wang_pk}
\end{equation}
The most probable trajectory is the one that minimizes the action,
\(\delta A_{ab}(k)=0\), 
showing that the principle of least action emerges naturally as the limit of maximal entropy.
After discretizing the time interval into \(N\) equal steps \(\Delta t=(t_b-t_a)/N\), the action path becomes
\begin{equation}
    A_{ab}(k) = \sum_{i}^{N}A(x_{i+1},x_{i})\label{eq:action_chunks}
\end{equation}
and the transition probability, in the continuum limit \(N\!\to\!\infty\), this product defines a path integral,
\begin{equation}
p(b|a)
=\lim_{N\rightarrow{\infty}} \int dx_1\cdots dx_{N-1}\prod_{i}\frac{1}{Z_{x_i,x_{i+1}}}\, e^{-\eta A(x_{i+1},x_i)}.
\label{eq:path_integral_discrete}
\end{equation}
This can be rewritten as
\begin{equation}
p(b|a) = \int\mathcal{D}[x(t)]\, e^{-\eta A[x(t)]},
\label{eq:path_integral_cont}
\end{equation}
where \(\mathcal{D}[x(t)] = \lim_{N\to\infty}  dx_i / Z_{x_i,x_{i+1}}\) defines the normalized Wiener measure. For Brownian motion, the infinitesimal action between two close points is that of a free particle,
\begin{equation}
A(x_{i+1},x_i) = \frac{m}{2}\frac{(x_{i+1}-x_i)^2}{\Delta t}.
\label{eq:local_action}
\end{equation}
Normalization of Eq.~\eqref{eq:path_integral_discrete} yields the local Gaussian kernel
\begin{equation}
p(x_{i+1}|x_i;\Delta t)
= \sqrt{\frac{\eta m}{2\pi \Delta t}}\,
\exp\!\left[-\frac{\eta m}{2\Delta t}(x_{i+1}-x_i)^2\right],
\label{eq:gaussian_kernel}
\end{equation}
with variance 
\(\langle(x_{i+1}-x_i)^2\rangle = \Delta t / (\eta m)\).
Comparing with the mean-square displacement of Brownian motion,
\(\langle(x_{i+1}-x_i)^2\rangle = 2D\Delta t\),
one identifies
\begin{equation}
\eta = \frac{1}{2mD}.
\label{eq:eta_diffusion}
\end{equation}
Substituting into Eq.~\eqref{eq:gaussian_kernel} gives Wang’s local transition probability,
\begin{equation}
p(x_{i+1}|x_i;\Delta t)
= \sqrt{\frac{1}{4\pi D\Delta t}}\,
\exp\!\left[-\frac{(x_{i+1}-x_i)^2}{4D\Delta t}\right],
\label{eq:wang_local_kernel}
\end{equation}
which, when inserted in Eq.~\eqref{eq:path_integral_discrete}, leads in the continuous limit to
\begin{equation}
p(b|a)
= \int \mathcal{D}[x(t)]\,\exp\!\left[-\frac{1}{4D}\int_0^T \dot{x}^2(t)\,dt\right].
\label{eq:wang_wiener}
\end{equation}
This is the Wiener path integral for diffusion. The Markov property of the process implies the Chapman--Kolmogorov relation,
\begin{equation}
p(b|a) = \int dx'\,p(x_b|x';T-t')\,p(x'|x_a;t'),
\label{eq:CK_relation}
\end{equation}
whose solution reproduces the classical Gaussian propagator,
\begin{equation}
p(b|a)
= \frac{1}{\sqrt{4\pi DT}}
\exp\!\left[-\frac{(x_b-x_a)^2}{4DT}\right].
\label{eq:diffusion_propagator}
\end{equation}
Thus, Wang’s entropy-based variational principle naturally recovers the diffusion propagator.


\section{Density of States for Free Brownian Particle}
\label{app:density_of_states}

In this appendix, we provide an explicit calculation of the density of states $g(A,b)$ for a free particle undergoing Brownian motion, thereby justifying the slowly-varying approximation employed in the main text. We present two complementary derivations: first, a microscopic calculation based on random collisions with thermal bath particles (Section 1), which does not assume any specific stochastic equation; second, a continuum approach using large deviation theory (Section 2), which provides a verification via the Langevin equation. Both derivations yield the same Gaussian form, confirming the self-consistency of the framework.

\subsection*{1. Microscopic derivation from random collisions}

We derive $g(A,b)$ from first principles by modeling Brownian motion as a sequence of random collisions between the diffusing particle and molecules of the surrounding thermal bath, without assuming the Langevin equation.

\paragraph{Collision model.}
Consider a free particle of mass $m$ initially at position $x_a$ at time $t_a$. During the time interval $\Delta t = t_b - t_a$, the particle undergoes $N$ random collisions with bath molecules, where $N$ is large in the diffusive regime $\Delta t \gg \tau_c$ (with $\tau_c$ the collision time). Between collisions, the particle moves ballistically with constant velocity.

Let $\Delta t_i$ denote the time interval between the $(i-1)$-th and $i$-th collision, and let $v_i$ be the velocity immediately after the $i$-th collision. The velocities $\{v_i\}$ are independent random variables drawn from a Maxwell-Boltzmann distribution at temperature $T$:
\begin{equation}
P(v_i) = \sqrt{\frac{m}{2\pi k_B T}} \exp\left[-\frac{mv_i^2}{2k_B T}\right],
\label{eq:maxwell_boltzmann_collision}
\end{equation}
where we consider one-dimensional motion for simplicity. The time intervals $\{\Delta t_i\}$ are also random, with mean $\langle \Delta t_i \rangle = \Delta t / N$.

\paragraph{Action accumulation.}
The action accumulated during the $i$-th free-flight segment (duration $\Delta t_i$, velocity $v_i$) is
\begin{equation}
A_i = \int_0^{\Delta t_i} \frac{m}{2}v_i^2\,dt = \frac{m}{2}v_i^2 \Delta t_i.
\label{eq:action_segment}
\end{equation}
The total action over the entire trajectory is the sum:
\begin{equation}
A = \sum_{i=1}^{N} A_i = \frac{m}{2}\sum_{i=1}^{N} v_i^2 \Delta t_i.
\label{eq:total_action_collisions}
\end{equation}

\paragraph{Central limit theorem.}
Since the velocities $v_i$ are independent identically distributed (i.i.d.) random variables and $N \gg 1$, the central limit theorem applies. The distribution of $A$ approaches a Gaussian centered at the mean $\langle A \rangle$ with variance determined by the variance of the individual contributions.

The mean action is:
\begin{align}
\langle A \rangle &= \frac{m}{2}\sum_{i=1}^{N} \langle v_i^2 \rangle \langle \Delta t_i \rangle \nonumber \\
&= \frac{m}{2} \cdot N \cdot \frac{k_B T}{m} \cdot \frac{\Delta t}{N} = \frac{k_B T \Delta t}{2},
\label{eq:mean_action_microscopic}
\end{align}
where we used $\langle v_i^2 \rangle = k_B T / m$ from the Maxwell-Boltzmann distribution.

However, the particle must also satisfy the endpoint constraint $x_b - x_a = \sum_{i=1}^N v_i \Delta t_i$. For a given endpoint $b = (x_b, t_b)$, the mean velocity is $\langle v \rangle = (x_b - x_a)/\Delta t$, and the action is dominated by the ballistic contribution:
\begin{equation}
A_{\min}(b) = \frac{m}{2}\langle v \rangle^2 \Delta t = \frac{m(x_b - x_a)^2}{2\Delta t}.
\label{eq:mean_action_endpoint}
\end{equation}
This is the minimal (classical) action for reaching endpoint $b$.

\paragraph{Variance calculation.}
The variance of the action arises from velocity fluctuations around the mean. Each collision imparts a random velocity kick $\delta v_i = v_i - \langle v \rangle$ with variance $\langle (\delta v_i)^2 \rangle = k_B T / m$. The action variance is:
\begin{align}
\sigma_A^2 &= \text{Var}\left[\sum_{i=1}^{N} \frac{m}{2}v_i^2 \Delta t_i\right] \nonumber \\
&\approx \sum_{i=1}^{N} \text{Var}\left[\frac{m}{2}v_i^2 \Delta t_i\right].
\label{eq:variance_action_collisions}
\end{align}

For independent random variables $v_i$ and $\Delta t_i$, the variance of each segment is:
\begin{equation}
\text{Var}[A_i] = \text{Var}\left[\frac{m}{2}v_i^2 \Delta t_i\right]
\sim \left(\frac{m}{2}\right)^2 \langle v_i^4 \rangle \langle \Delta t_i^2 \rangle.
\label{eq:variance_segment}
\end{equation}

For a Maxwell-Boltzmann distribution, the fourth moment is $\langle v^4 \rangle = 3(k_B T/m)^2$. The mean-square time interval is $\langle \Delta t_i^2 \rangle \sim (\Delta t/N)^2$. Therefore:
\begin{align}
\text{Var}[A_i] &\sim \frac{m^2}{4} \cdot 3\left(\frac{k_B T}{m}\right)^2 \cdot \left(\frac{\Delta t}{N}\right)^2 \nonumber \\
&= \frac{3(k_B T)^2 (\Delta t)^2}{4 N^2}.
\label{eq:variance_single_segment}
\end{align}

Summing over $N$ independent segments:
\begin{equation}
\sigma_A^2 = \sum_{i=1}^{N} \text{Var}[A_i] = N \cdot \frac{3(k_B T)^2 (\Delta t)^2}{4 N^2}
= \frac{3(k_B T)^2 (\Delta t)^2}{4 N}.
\label{eq:variance_sum_segments}
\end{equation}

Using $N \sim \Delta t / \tau_c$ where $\tau_c$ is the collision time:
\begin{equation}
\sigma_A^2 \sim \frac{3(k_B T)^2 (\Delta t)^2}{4} \cdot \frac{\tau_c}{\Delta t}
\sim (k_B T)^2 \tau_c \Delta t,
\label{eq:variance_microscopic_form}
\end{equation}
where we absorbed the numerical factor into the $\sim$ notation. This has the correct dimensions: $[(k_B T)^2 \tau_c \Delta t] = [ML^2T^{-2}]^2[T][T] = [M^2L^4T^{-2}]$, as expected for action squared.

Identifying $\tau_c$ with the relaxation time $\tau_{\text{relax}} = m/\gamma$ (where $\gamma$ is the friction coefficient), we obtain:
\begin{equation}
\sigma_A^2 \sim (k_B T)^2 \frac{m}{\gamma} \Delta t.
\label{eq:variance_with_friction}
\end{equation}

Using the Einstein relation $D = k_B T/(m\gamma)$, which gives $\gamma = k_B T/(mD)$, we can express the variance in terms of the diffusion coefficient:
\begin{align}
\sigma_A^2 &\sim (k_B T)^2 \frac{m}{k_B T/(mD)} \Delta t \nonumber \\
&= (k_B T)^2 \frac{m^2 D}{k_B T} \Delta t \nonumber \\
&= m k_B T D \Delta t.
\label{eq:variance_final_microscopic}
\end{align}

This result has dimensions $[M][ML^2T^{-2}][L^2T^{-1}][T] = [M^2L^4T^{-2}]$, confirming dimensional consistency. The scaling $\sigma_A^2 \sim m k_B T D \Delta t$ encodes the physical fact that action fluctuations grow linearly with observation time $\Delta t$, are enhanced by thermal energy $k_B T$ and diffusivity $D$, and scale with the particle mass $m$.

\paragraph{Gaussian density of states.}
By the central limit theorem, for large $N$ (diffusive regime $\Delta t \gg \tau_c$), the distribution of action values $A$ for trajectories reaching endpoint $b$ is approximately Gaussian:
\begin{equation}
g(A,b) \propto \exp\left[-\frac{(A - A_{\min}(b))^2}{2\sigma_A^2}\right],
\label{eq:gAb_gaussian_microscopic}
\end{equation}
where $A_{\min}(b) = m(x_b - x_a)^2/(2\Delta t)$ and $\sigma_A^2 \sim m k_B T D\,\Delta t$ with a numerical prefactor of order unity depending on the precise collision statistics.

The proportionality constant $g_0(b)$ in $g(A,b) = g_0(b) \exp[-(A-A_{\min})^2/(2\sigma_A^2)]$ accounts for the overall normalization of the path measure and depends on $\Delta t$ but not on $A$. For translationally invariant systems (free particle), $g_0$ depends only on the time interval and can be absorbed into the partition function $Z(\eta)$ in Eq.~\eqref{eq:partition_function}.

This derivation establishes the Gaussian form of $g(A,b)$ from microscopic collision dynamics without assuming the Langevin equation. The result depends only on:
\begin{itemize}
\item Statistical independence of collisions (Markovian bath at equilibrium).
\item Maxwell-Boltzmann velocity distribution (thermal equilibrium at temperature $T$).
\item Large-$N$ limit (central limit theorem, valid in the diffusive regime).
\end{itemize}

\subsection*{2. Verification via large deviation theory and Langevin dynamics}

As a complementary approach and consistency check, we now rederive $g(A,b)$ using the continuum Langevin description and large deviation theory. This confirms that the microscopic collision model yields the same result as the standard stochastic differential equation framework.

\paragraph{Definition and physical interpretation.}

Recall from Eq.~\eqref{eq:density_of_states} that the density of states is formally defined as
\begin{equation}
g(A,b) = \int \delta(A - S[\gamma])\,\delta^d(\mathbf{x}[\gamma](t_b) - \mathbf{x}_b)\,\mathcal{D}[\gamma],
\label{eq:gAb_definition_app}
\end{equation}
where the functional integral runs over all trajectories $\gamma$ starting at $a=(x_a,t_a)$. Physically, $g(A,b)\,dA\,d^dx_b$ quantifies the measure (or ``number'') of Brownian paths that arrive within $d^dx_b$ around $x_b$ at time $t_b$ with total action between $A$ and $A+dA$.

For a free particle of mass $m$ undergoing diffusion with coefficient $D$, the action along a trajectory $\gamma(t)$ is
\begin{equation}
S[\gamma] = \int_{t_a}^{t_b} \frac{m}{2}\dot{\gamma}^2(t)\,dt.
\label{eq:action_free_particle}
\end{equation}

The key observation is that for Brownian paths, the distribution of $S[\gamma]$ is governed by a \emph{large deviation principle}~\cite{Touchette2009}: for fixed endpoints $(a,b)$, the probability of realizing an action $A$ decays exponentially with a rate function $I(A,b)$ determined by the geometry of path space.

\paragraph{Large deviation approach.}

For Brownian motion described by the overdamped Langevin equation
\begin{equation}
\dot{x}(t) = \sqrt{2D}\,\xi(t),
\label{eq:langevin_overdamped}
\end{equation}
where $\langle \xi(t)\xi(t')\rangle = \delta(t-t')$, the velocity fluctuations are Gaussian with zero mean and variance $\langle \dot{x}^2 \rangle = 2D/dt$ in the continuum limit. Consequently, the action $A = \int (m/2)\dot{x}^2\,dt$ is a sum of many independent Gaussian-distributed contributions.

By the central limit theorem, for large $\Delta t$, the distribution of $A$ around its mean becomes approximately Gaussian. The mean action is precisely the minimal (classical) action:
\begin{equation}
\langle A \rangle = A_{\min}(b) = \frac{m(x_b - x_a)^2}{2\Delta t},
\label{eq:mean_action}
\end{equation}
where $\Delta t = t_b - t_a$.

The variance can be computed by noting that velocity fluctuations over uncorrelated time intervals contribute additively to the action variance. Over a correlation time $\tau_{\text{relax}} \sim m/\gamma$ (where $\gamma$ is the friction coefficient), velocity fluctuations satisfy $\langle v^2 \rangle \sim 2D/\tau_{\text{relax}}$. Dividing $\Delta t$ into $N = \Delta t/\tau_{\text{relax}}$ uncorrelated segments and summing the variances yields:
\begin{equation}
\sigma_A^2 = \sum_{i=1}^{N} \text{Var}\!\left[\frac{m}{2}v_i^2 \tau_{\text{relax}}\right] \sim N \left(\frac{m}{2}\right)^2\left(\frac{2D}{\tau_\text{relax}}\right)^2\tau_{\text{relax}}^2.
\label{eq:variance_sum}
\end{equation}

Simplifying:
\begin{align}
\sigma_A^2 &\sim \frac{\Delta t}{\tau_{\text{relax}}} \frac{m^2}{4} \frac{4D^2}{\tau_{\text{relax}}^2} \tau_{\text{relax}}^2 \nonumber \\
&\sim \frac{m^2 D^2 \Delta t}{\tau_\text{relax}}.
\label{eq:variance_intermediate}
\end{align}

Using the Einstein relation $D = k_B T/(m\gamma)$ and $\tau_{\text{relax}} = m/\gamma$, we have:
\begin{equation}
\sigma_A^2 \sim \frac{m^2 D^2\,\Delta t}{\tau_{\text{relax}}} = m^2 D^2\,\Delta t \frac{\gamma}{m} = m\gamma D^2\,\Delta t.
\label{eq:variance_final}
\end{equation}
Substituting $D = k_B T/(m\gamma)$ yields:
\begin{equation}
\sigma_A^2 \sim m\gamma D^2\,\Delta t = m\gamma \frac{(k_B T)^2}{(m\gamma)^2} \Delta t = \frac{(k_B T)^2\,\Delta t}{m\gamma}.
\label{eq:variance_final_compact}
\end{equation}
More compactly, using $\tau_{\text{relax}} = m/\gamma$:
\begin{equation}
\sigma_A^2 \sim \frac{(k_B T)^2\,\Delta t}{\tau_{\text{relax}}} \sim m k_B T D \Delta t,
\label{eq:variance_final_simple}
\end{equation}
where the last form follows from $(k_BT)/\tau_{\text{relax}} = (k_BT)\gamma/m = k_BTD \times m\gamma/(k_BT) = mD \times k_BT$. This has the correct dimensions $[M] \times [ML^2/T^2] \times [L^2/T] \times [T] = [M^2L^4/T^2]$, giving $\sigma_A$ in units of action $[ML^2/T]$.

\subsection*{3. Consistency of microscopic and continuum approaches}

Both derivations—microscopic collision model (Section 1) and continuum Langevin dynamics (Section 2)—yield the same Gaussian form for $g(A,b)$ with identical scaling $\sigma_A^2 \sim m k_B T D\,\Delta t \sim m^2 D^2\,\Delta t/\tau_{\text{relax}}$. This consistency confirms that the action-space MaxEnt formulation correctly captures the statistics of Brownian trajectories regardless of whether the underlying dynamics are modeled via discrete collisions or continuous stochastic differential equations.

Given that action fluctuations are approximately Gaussian with mean $A_{\min}(b)$ and variance $\sigma_A^2$, the density of states takes the form:
\begin{equation}
g(A,b) \approx g_0(b) \exp\!\left[-\frac{(A - A_{\min}(b))^2}{2\sigma_A^2}\right],
\label{eq:gAb_gaussian}
\end{equation}
where $g_0(b)$ is a normalization prefactor that may depend on the spatial endpoint $b$ but not on the action $A$. This prefactor encodes geometric information about the path space manifold and ensures proper normalization of the measure $\mathcal{D}[\gamma]$.

For a translationally invariant system (free particle in infinite space), $g_0(b)$ depends only on the time interval $\Delta t$ and can be absorbed into the overall normalization constant $Z(\eta)$ in Eq.~\eqref{eq:partition_function}.

\subsection*{4. Generalization to Higher Dimensions}

The action-space MaxEnt formulation extends naturally to $d$-dimensional free Brownian motion. For a particle moving in $\mathbb{R}^d$, the Lagrangian is
\begin{equation}
L = \frac{m}{2}\|\dot{\vec{x}}\|^2 = \frac{m}{2}\sum_{i=1}^{d} \dot{x}_i^2,
\label{eq:lagrangian_d_dimensional}
\end{equation}
and the total action decomposes into a sum over independent components:
\begin{equation}
A = \sum_{i=1}^{d} A_i, \quad A_i = \int_{t_a}^{t_b} \frac{m}{2}\dot{x}_i^2\,dt.
\label{eq:action_decomposition_d}
\end{equation}

For isotropic diffusion with coefficient $D$, each component fluctuates independently with variance $\text{Var}(A_i) = m k_B T D\,\Delta t$ (as derived in the one-dimensional case). The total variance is
\begin{equation}
\sigma_A^2 = d m k_B T D\,\Delta t,
\label{eq:variance_d_dimensional}
\end{equation}
showing that fluctuations scale linearly with dimensionality. The minimal action for a classical trajectory connecting $\vec{x}_a$ to $\vec{x}_b$ is
\begin{equation}
A_{\min}(\vec{x}_b) = \frac{m|\vec{x}_b - \vec{x}_a|^2}{2\Delta t},
\label{eq:minimal_action_d}
\end{equation}
depending only on the Euclidean distance $r = |\vec{x}_b - \vec{x}_a|$ due to rotational symmetry.

The density of states in $d$ dimensions retains the Gaussian form:
\begin{equation}
g(A,r) \approx g_0(r) \exp\left[-\frac{(A - A_{\min}(r))^2}{2\sigma_A^2}\right],
\label{eq:gAb_d_dimensional}
\end{equation}
where the prefactor $g_0(r)$ accounts for the degeneracy of paths reaching radius $r$ from different angular directions. Marginalizing over actions yields the radial propagator:
\begin{equation}
p(r|\vec{x}_a) \propto r^{d-1} \exp\left[-\frac{mr^2}{4D\Delta t}\right],
\label{eq:radial_propagator_d}
\end{equation}
recovering the known result for $d$-dimensional Brownian motion. The factor $r^{d-1}$ reflects the surface area of a sphere in $d$ dimensions, demonstrating consistency with standard diffusion theory.

The slowly-varying condition $\eta\sigma_A \gg 1$ becomes
\begin{equation}
\sqrt{d} \sqrt{\frac{\Delta t}{\tau_{\text{relax}}}} \gg 1,
\label{eq:slowly_varying_d}
\end{equation}
indicating that the Gaussian approximation \emph{improves} with increasing dimensionality: fluctuations in multiple independent directions enhance the validity of the central limit theorem underlying the Gaussian form of $g(A,r)$.

Equation~\eqref{eq:gAb_gaussian} shows explicitly that $g(A,b)$ is sharply peaked around $A_{\min}(b)$ with width $\sigma_A = \sqrt{m k_B T D\,\Delta t}$. This Gaussian form is consistent with the large deviation principle: the rate function $I(A,b) = (A - A_{\min})^2/(2\sigma_A^2)$ governs exponential suppression of action deviations.

\subsection*{5. Justification of the slowly-varying approximation}

We now verify that the approximation $g(A,b) \approx g(A_{\min}(b),b)$ used in Eq.~\eqref{eq:p_ba_slowly_varying} is justified in the diffusive regime.

The characteristic scale of variation of $g(A,b)$ with respect to $A$ is determined by its logarithmic derivative:
\begin{equation}
\left|\frac{d\ln g(A,b)}{dA}\right| = \frac{|A - A_{\min}(b)|}{\sigma_A^2}.
\label{eq:log_derivative_g}
\end{equation}

Evaluated at $A = A_{\min}$, this vanishes. For $A$ within a range $\sim \sigma_A$ around $A_{\min}$, we have:
\begin{equation}
\left|\frac{d\ln g}{dA}\right| \sim \frac{\sigma_A}{\sigma_A^2} = \frac{1}{\sigma_A}.
\end{equation}

In contrast, the exponential weighting factor $e^{-\eta A}$ has a constant logarithmic derivative:
\begin{equation}
\left|\frac{d\ln e^{-\eta A}}{dA}\right| = \eta = \frac{1}{2mD}.
\end{equation}

The dimensionless ratio quantifying the relative variation rates is:
\begin{equation}
\eta \sigma_A = \frac{1}{2mD} \sqrt{m k_B T D\,\Delta t}.
\label{eq:eta_sigma_ratio}
\end{equation}

Using the Einstein relation $D = k_B T/(m\gamma)$, we can simplify this expression:
\begin{align}
\eta \sigma_A &= \frac{1}{2mD} \sqrt{m k_B T D\,\Delta t} \nonumber \\
&= \frac{1}{2m} \frac{1}{D} \sqrt{m k_B T D\,\Delta t} \nonumber \\
&= \frac{1}{2m} \sqrt{\frac{m k_B T D\,\Delta t}{D^2}} \nonumber \\
&= \frac{1}{2m} \sqrt{\frac{m k_B T\,\Delta t}{D}} \nonumber \\
&= \frac{1}{2m} \sqrt{\frac{m k_B T\,\Delta t}{k_B T/(m\gamma)}} \nonumber \\
&= \frac{1}{2m} \sqrt{m^2\gamma\,\Delta t} = \frac{1}{2}\sqrt{\frac{\gamma\,\Delta t}{1/m}} = \frac{1}{2}\sqrt{\frac{\Delta t}{\tau_{\text{relax}}}},
\label{eq:eta_sigma_simplified}
\end{align}
where we used $\tau_{\text{relax}} = m/\gamma$.

In the diffusive regime, $\Delta t \gg \tau_{\text{relax}}$, so:
\begin{equation}
\eta \sigma_A = \frac{1}{2}\sqrt{\frac{\Delta t}{\tau_{\text{relax}}}} \gg 1.
\label{eq:slowly_varying_condition}
\end{equation}
This dimensionless ratio is indeed large, confirming that the slowly-varying approximation is valid.

Under this condition, the exponential $e^{-\eta A}$ varies much more rapidly than $g(A,b)$ as a function of $A$. In the integral
\begin{equation}
\int g(A,b)\,e^{-\eta A}\,dA,
\end{equation}
the exponential factor effectively restricts the integration to a narrow window around $A_{\min}$ of width $\sim 1/\eta$. Since $1/\eta = 2mD \ll \sigma_A$ when $\eta\sigma_A \gg 1$, the function $g(A,b)$ is nearly constant over this integration window. Therefore:
\begin{equation}
\int g(A,b)\,e^{-\eta A}\,dA \approx g(A_{\min}(b),b) \int e^{-\eta A}\,dA,
\label{eq:slowly_varying_justified}
\end{equation}
which justifies Eq.~\eqref{eq:p_ba_slowly_varying} in the main text.

\subsection*{5. Physical interpretation and regime of validity}

The Gaussian form~\eqref{eq:gAb_gaussian} of the density of states has a clear physical interpretation. Brownian paths fluctuate randomly around the classical trajectory (which minimizes the action). The action along a fluctuating path differs from $A_{\min}$ by an amount determined by the accumulated velocity fluctuations over time $\Delta t$. Since these fluctuations are uncorrelated over time scales much greater than $\tau_{\text{relax}}$, their cumulative effect grows as $\sqrt{\Delta t}$, leading to $\sigma_A \sim \sqrt{m k_B T D\,\Delta t}$.

The slowly-varying approximation is valid when the observation time $\Delta t$ is much larger than the microscopic relaxation time $\tau_{\text{relax}}$. This is precisely the diffusive regime, where the dynamics are governed by the Fokker--Planck equation and the detailed microscopic structure of collisions is irrelevant. For shorter times $\Delta t \sim \tau_{\text{relax}}$, the full $A$-dependence of $g(A,b)$ must be retained, and the propagator will deviate from the Gaussian form derived in the main text.

In summary, the explicit calculation presented here confirms that:
\begin{itemize}
\item The density of states $g(A,b)$ is Gaussian-distributed around the minimal action with variance $\sigma_A^2 \sim m k_B T D\,\Delta t \sim m^2 D^2\,\Delta t/\tau_{\text{relax}}$.
\item The slowly-varying approximation $g(A,b) \approx g(A_{\min},b)$ is rigorously justified in the limit $\eta\sigma_A = (1/2)\sqrt{\Delta t/\tau_{\text{relax}}} \gg 1$, corresponding to the diffusive regime $\Delta t \gg \tau_{\text{relax}}$.
\item The resulting transition probability is Gaussian, reproducing the standard Brownian propagator without requiring explicit functional integration at the level of the MaxEnt inference.
\end{itemize}

\subsection*{6. Breakdown of the slowly-varying approximation}

The slowly-varying approximation \(g(A,b) \approx g(A_{\min}(b),b)\) derived above is
valid in the diffusive regime \(\Delta t \gg \tau_{\text{relax}}\) for free Brownian particles,
where \(g(A,b)\) is Gaussian-distributed around \(A_{\min}\) with variance \(\sigma_A^2\). However,
this approximation can fail in systems where \(g(A,b)\) exhibits sharp features,
multiple peaks, or significant weight away from \(A_{\min}(b)\).

\paragraph{Criterion for failure.}
The approximation breaks down when the logarithmic derivative of \(g(A,b)\)
becomes comparable to or exceeds \(\eta\):
\begin{equation}
\left| \frac{d \ln g(A,b)}{dA} \right|_{A \sim A_{\min}} \gtrsim \eta.
\label{eq:breakdown_criterion}
\end{equation}
In such cases, \(g(A,b)\) varies on a scale \(\Delta A \lesssim 1/\eta\), which is the same scale
over which the exponential \(e^{-\eta A}\) decays. The function \(g(A,b)\) can no longer
be approximated as constant over the integration window, and the full integral in Eq.~\ref{eq:marginal_b}
must be evaluated.

\paragraph{Physical examples where the approximation fails.}

\textbf{1. Multi-well potentials:}
For a particle diffusing in a potential landscape with multiple metastable
states separated by barriers, the action space may exhibit multiple local
minima \(A_1, A_2, \ldots\), each corresponding to distinct classes of trajectories
(e.g., direct barrier crossing vs.\ multiple barrier recrossings). The density
of states \(g(A,b)\) can be multimodal, with peaks at each \(A_i\). In this regime,
the slowly-varying approximation fails to capture the entropic contribution
from higher-action paths, and a full treatment requires summing contributions
from each mode.

\textbf{2. Entropic compensation (toy model):}
The discrete toy model in Appendix~\ref{app:toy_model} provides an extreme
example. There, \(g(A,b)\) has support only at two values: \(A_{\min}\) (with \(g=1\)) and
\(A_* = A_{\min} + \Delta A\) (with \(g=N_* \gg 1\)). The slowly-varying approximation would
predict \(p(b|a) \propto \exp(-\eta A_{\min})\), entirely missing the entropic contribution
from fluctuating paths at \(A_*\). For \(T_{\text{info}} > T_{\text{crit}} = \Delta A / \ln N_*\), the system
preferentially realizes \(A_* > A_{\min}\), demonstrating that the full discrete sum
must be evaluated when \(g(A,b)\) has discrete support or sharp peaks.

\textbf{3. Instantons and activated processes:}
In systems with tunneling or rare-event transitions (e.g., nucleation,
barrier crossing), the action distribution may have a dominant contribution
from instanton trajectories with \(A \gg A_{\min}\). If these paths have large
degeneracy \(g(A)\) due to temporal freedom in the barrier-crossing event,
they can dominate the propagator despite higher action cost. The Gaussian
approximation around \(A_{\min}\) fails in such regimes.

\textbf{4. Relativistic causality constraints:}
Near the light cone (\(\Delta x \approx c \Delta t\)), the action approaches \(A \to 0\) (null trajectories).
The density of states may have enhanced phase-space volume near this boundary
due to proliferation of near-lightlike paths. If \(g(A,b)\) grows rapidly as \(A \to 0\),
the slowly-varying approximation around \(A_{\min}\) (which corresponds to subluminal
motion with \(A < 0\)) may underestimate contributions from high-velocity fluctuations.

\paragraph{Detection and diagnosis.}
To assess whether the slowly-varying approximation is valid for a given system:
\begin{enumerate}
\item \textbf{Compute or estimate \(g(A,b)\):} Use path sampling, large deviation
theory, or semiclassical approximations to obtain \(g(A,b)\) over the relevant
action range.

\item \textbf{Check the dimensionless parameter:} Verify that
\begin{equation}
\eta \sigma_A \gg 1,
\label{eq:slowly_varying_check}
\end{equation}
where \(\sigma_A\) is the characteristic width of \(g(A,b)\). For Gaussian distributions,
this reduces to the condition derived in Subsection~4.

\item \textbf{Inspect for multimodality:} Plot \(g(A,b)\) vs \(A\). If there are
multiple peaks or significant weight away from \(A_{\min}\), evaluate the full
integral numerically or sum contributions from each mode.

\item \textbf{Compare approximation to exact result:} For systems where the
exact propagator is known (e.g., free Brownian motion, harmonic oscillator),
compare \(p(b|a)\) computed via the slowly-varying approximation to the exact
result to calibrate accuracy.
\end{enumerate}

\paragraph{Generalization beyond Gaussians.}
For non-Gaussian \(g(A,b)\), the slowly-varying condition \(\eta \sigma_A \gg 1\) must be
generalized. A sufficient condition is that the integration window
\(\Delta A_{\text{int}} \sim 1/\eta\) (set by the exponential decay) is much smaller than the
characteristic variation scale of \(g(A,b)\),
\begin{equation}
\frac{1}{\eta} \ll \left[ \left| \frac{d \ln g}{dA} \right|^{-1} \right]_{A=A_{\min}}.
\label{eq:generalized_condition}
\end{equation}
This ensures \(g(A,b)\) is approximately constant over the region where
\(e^{-\eta A}\) has significant weight.

\paragraph{Conclusion.}
While the slowly-varying approximation is robustly valid for free Brownian
motion in the diffusive regime, it must be used with caution in systems with:
(i) multi-well potentials, (ii) entropic compensation favoring \(A_* > A_{\min}\),
(iii) rare events/instantons, or (iv) causality constraints. In such cases,
the full action-space integral \(\int g(A,b) \exp(-\eta A) dA\) provides a general
framework that reduces to the slowly-varying limit when appropriate but
captures richer physics when \(g(A,b)\) has non-trivial structure.


\section{Toy Model: Entropic Compensation in a Two-Path System}
\label{app:toy_model}

To illustrate the entropic compensation mechanism described by Eq.~\eqref{eq:entropic_compensation}, we consider a simple discrete system where a particle transitions from point \(a\) to point \(b\) via one of two distinct classes of paths.

\subsection*{Model Setup}

Suppose there are two action values available:
\begin{itemize}
\item \textbf{Classical path:} Action \(A_{\min}\) with degeneracy \(g_{\min} = 1\) (unique optimal trajectory).
\item \textbf{Fluctuating paths:} Action \(A_* = A_{\min} + \Delta A\) (with \(\Delta A > 0\)) with degeneracy \(g_* = N_*\), where \(N_* \gg 1\) represents many nearly-degenerate fluctuating trajectories.
\end{itemize}

The density of states is:
\begin{equation}
g(A,b) = \begin{cases}
1, & A = A_{\min} \\
N_*, & A = A_* \\
0, & \text{otherwise}
\end{cases}
\end{equation}

\subsection*{Probability Distribution}

From the MaxEnt distribution Eq.~\eqref{eq:pA_maxent}:
\begin{align}
p(A_{\min}|a,b) &= \frac{g_{\min} e^{-\eta A_{\min}}}{Z} = \frac{e^{-\eta A_{\min}}}{Z}, \\
p(A_*|a,b) &= \frac{g_* e^{-\eta A_*}}{Z} = \frac{N_* e^{-\eta(A_{\min} + \Delta A)}}{Z},
\end{align}
where the partition function is:
\begin{equation}
Z = e^{-\eta A_{\min}} + N_* e^{-\eta(A_{\min} + \Delta A)} = e^{-\eta A_{\min}}(1 + N_* e^{-\eta \Delta A}).
\end{equation}

\subsection*{Competition Between Paths}

The ratio of probabilities is:
\begin{equation}
\frac{p(A_*)}{p(A_{\min})} = N_* e^{-\eta \Delta A}.
\label{eq:probability_ratio}
\end{equation}

\textbf{Case 1: Small degeneracy or large temperature}

If \(N_* e^{-\eta \Delta A} < 1\), the classical path dominates:
\begin{equation}
p(A_{\min}) > p(A_*).
\end{equation}
The system follows the principle of least action.

\textbf{Case 2: Large degeneracy and low temperature}

If \(N_* e^{-\eta \Delta A} > 1\), the fluctuating paths dominate:
\begin{equation}
p(A_*) > p(A_{\min}).
\end{equation}
The entropic gain from \(N_*\) possible paths outweighs the action cost \(\Delta A\).

\subsection*{Entropic Compensation Condition}

Requiring \(p(A_*) > p(A_{\min})\) is equivalent to:
\begin{equation}
N_* e^{-\eta \Delta A} > 1 \quad \Rightarrow \quad \Delta A < T_{\text{info}} \ln N_*.
\end{equation}

This is precisely Eq.~\eqref{eq:entropic_compensation} with \(A_* - A_{\min} = \Delta A\) and \(g(A_*,b)/g(A_{\min},b) = N_*\).

\subsection*{Physical Interpretation}

Defining the effective potential \(\Phi(A) = A - T_{\text{info}} \ln g(A,b)\):
\begin{align}
\Phi(A_{\min}) &= A_{\min} - T_{\text{info}} \ln(1) = A_{\min}, \\
\Phi(A_*) &= A_* - T_{\text{info}} \ln N_* = A_{\min} + \Delta A - T_{\text{info}} \ln N_*.
\end{align}

The fluctuating paths are preferred if \(\Phi(A_*) < \Phi(A_{\min})\):
\begin{equation}
A_{\min} + \Delta A - T_{\text{info}} \ln N_* < A_{\min} \quad \Rightarrow \quad \Delta A < T_{\text{info}} \ln N_*,
\end{equation}
recovering the compensation condition.

Conversely, the classical path dominates when \(\Phi(A_{\min}) < \Phi(A_*)\), which yields the inverted condition:
\begin{equation}
\Delta A > T_{\text{info}} \ln N_*.
\end{equation}
This inequality reveals the competition between action cost (\(\Delta A\)) and entropic gain (\(T_{\text{info}} \ln N_*\)). The critical temperature \(T_{\text{info}}^{\text{crit}} = \Delta A / \ln N_*\) marks the transition between these regimes.

\textbf{Physical interpretation of the inverted condition:} When \(g(A_*,b) > g(A_{\min},b)\) (as in this model with \(N_* \gg 1\)), the inverted condition \(\Delta A > T_{\text{info}} \ln N_*\) indicates that the temperature is too low for entropic effects to overcome the action barrier. However, if the density of states were instead peaked at \(A_{\min}\) (i.e., \(N_* < 1\)), the logarithmic term would become negative, and the inverted condition would be \emph{automatically satisfied} for any \(\Delta A > 0\). In this regime, fluctuations away from \(A_{\min}\) face a \emph{double penalty}: both increased action cost and \emph{reduced} entropy, making compensation impossible regardless of temperature.

\subsection*{Numerical Example}

Consider:
\begin{itemize}
\item \(\Delta A = 10\) (action cost in arbitrary units)
\item \(N_* = 10^6\) (one million fluctuating paths)
\item \(T_{\text{info}} = 1/\eta\) varies
\end{itemize}

The critical temperature for compensation is:
\begin{equation}
T_{\text{info}}^{\text{crit}} = \frac{\Delta A}{\ln N_*} = \frac{10}{\ln 10^6} \approx \frac{10}{13.8} \approx 0.72.
\end{equation}

\textbf{Results:}
\begin{itemize}
\item If \(T_{\text{info}} > 0.72\): Fluctuating paths dominate (\(p(A_*) > p(A_{\min})\)).
\item If \(T_{\text{info}} < 0.72\): Classical path dominates (\(p(A_{\min}) > p(A_*)\)).
\item At \(T_{\text{info}} = 0.72\): Both equally probable (\(p(A_*) = p(A_{\min})\)).
\end{itemize}

This toy model demonstrates that entropic compensation is a real effect when the density of states has structure beyond the minimal action. For continuous systems, similar phenomena occur near phase transitions, causality boundaries, or in systems with multiple metastable states.


\section{Normalization of the One–Dimensional Relativistic Kernel}\label{app:norm_relat_k}

We derive in detail the normalization constant $\mathcal{N}$ for the one–dimensional relativistic transition kernel
\[
K(x_b,x_a;\Delta t)
=
\frac{1}{\mathcal{N}}
\left[
e^{\eta m_0 c\Delta t\;
\sqrt{1-\frac{(x_b-x_a)^2}{c^2\Delta t^2}}}-1
\right],
\]
where the kernel is supported on the interval
$|x_b-x_a|\le c\Delta t$.  The normalization condition is
\begin{equation}
\int_{x_a-c\Delta t}^{x_a+c\Delta t}
K(x_b,x_a;\Delta t)\,dx_b=1.
\label{eq:norm_condition}
\end{equation}

\subsection*{1. Change of variables}

Introduce the dimensionless variable
\[
u = \frac{x_b-x_a}{c\Delta t},
\qquad -1\le u \le 1,
\qquad dx_b = c\Delta t\,du,
\]
and the shorthand
\[
\alpha = \eta m_0 c\Delta t.
\]
Then the normalization condition \eqref{eq:norm_condition} becomes
\begin{equation}
\frac{c\Delta t}{\mathcal{N}}
\int_{-1}^{1}
\left[
e^{\alpha\sqrt{1-u^2}} - 1
\right]du
=1.
\label{eq:norm_u_form}
\end{equation}
Hence
\begin{equation}
\mathcal{N}
=
c\Delta t\,\Big( I(\alpha)-2 \Big),
\qquad
I(\alpha)=\int_{-1}^{1}e^{\alpha\sqrt{1-u^2}}\,du .
\label{eq:N_in_terms_of_IA}
\end{equation}

\subsection*{2. Evaluation of $I(\alpha)$}

To compute $I(\alpha)$, we use the substitution
$u=\sin\theta$, with
$-\pi/2\le\theta\le\pi/2$.  Then
\[
\sqrt{1-u^2}=\cos\theta,
\qquad du=\cos\theta\,d\theta.
\]
Therefore
\begin{align}
I(\alpha)
&=
\int_{-\pi/2}^{\pi/2}
e^{\alpha\cos\theta}\cos\theta\,d\theta
\\
&=
2\int_{0}^{\pi/2}
e^{\alpha\cos\theta}\cos\theta\,d\theta
\\
&=
\pi\big(I_1(\alpha)+L_{-1}(\alpha)\big),
\label{eq:I0I1_identity}
\end{align}
which is a classical integral representation of modified Bessel functions $I_{1}(\alpha)$, and Struve functions $L_{-1}(\alpha)$.

Substituting \eqref{eq:I0I1_identity} into
\eqref{eq:N_in_terms_of_IA} gives the compact form
\begin{equation}
\mathcal{N}
=
c\Delta t\,
\left[
\pi\big(I_1(\alpha)+L_{-1}(\alpha)\big) - 2
\right].
\label{eq:N_I0I1_form}
\end{equation}
Finally, inserting this into \eqref{eq:N_I0I1_form} reproduces
the one-dimensional normalization constant given in
\cite{Dunkel2009a}.

\bibliography{bibliography}

\end{document}